\documentclass{appolb}
\usepackage{epsfig,amssymb,xspace}
     \newcommand{\pathnow}{}
\font\twelvegoth=eufm10 at 14.4pt
\newfam\gothfam
\textfont\gothfam=\twelvegoth

\font\twelveron=eusm10 at 14.4pt
\newfam\ronfam
\textfont\ronfam=\twelveron

\newcommand{\nc}{\newcommand}
\nc{\ds}{\displaystyle}        \nc{\ts}{\textstyle}
\nc{\rf}[1]{Fig.\,\ref{#1}}    \nc{\rt}[1]{table\,\ref{#1}}
\nc{\req}[1]{Eq.\,(\ref{#1})}  \nc{\eps}{\varepsilon}
\nc{\beq}{\begin{equation}}     \nc{\beql}[1]{\begin{equation}\label{#1}}
\nc{\eeq}{\end{equation}}        
\nc{\beqa}{\begin{eqnarray}}   \nc{\eeqa}{\end{eqnarray}}       
\nc{\bfi}{\begin{figure}}       \nc{\efi}{\end{figure}}

\nc{\clbox}[1]{

\begin{center}\framebox{#1}\end{center}
}
\nc{\flbox}[1]{\fboxrule=5pt\framebox{#1}\fboxrule=2pt}
\begin{document}
\title{Strangeness and Statistical Hadronization: 
\\How to Study  Quark--Gluon Plasma
}
\author{Jan Rafelski
\address{Department of Physics, University of Arizona, Tucson, AZ85721}
\and
Jean Letessier
\address{Laboratoire de Physique 
                Th\'eorique et Hautes Energies\\
     Universit\'e Paris 7, 2 place Jussieu, F--75251 Cedex 05.}
}
\maketitle
\begin{abstract}
Statistical hadronization is presented as mechanism for (strange) particle 
production from a deconfined  quark--gluon plasma (QGP) fireball.  We first 
consider hadronic resonance production at RHIC as a test of the model.
 We present in detail how the hadrochemistry determines 
particle multiplicities and in case of sudden hadronization 
allows investigation of QGP  properties.
A comparative study of strange hadron 
production at  SPS and RHIC is presented. The energy dependence
of physical observables shows regularities and a potential 
discontinuity in the low RHIC range, when comparing  these different energy 
domains. Considering  the 
energy scan program  at CERN-SPS we show that the $K^+/\pi^+$ 
discontinuity is a baryon density effect.
\end{abstract}

\section{Introduction}
\subsection{Setting the stage}\label{stage}
It is believed today that a new  state of matter has been formed 
in relativistic nuclear collisions at BNL-RHIC and  at CERN-SPS,
and perhaps even in 1800 GeV  $p\bar p$ elementary interactions. 
The question of considerable interest is if  this is 
the hot quark matter (quark--gluon
plasma, QGP) state of matter. The paradigm of QGP 
 originates in  the  quantum many body theory of 
quark matter \cite{Car73,ILL74}, which lead on to the
formal recognition within the framework of asymptotically free 
quantum-chromodynamics (QCD) that at a very high temperature 
perturbative quark matter state must exist \cite{Col75}. 

Further reasoning based on a study of the `boiling state' of dense
hadron gas within the scheme of Hagedorn's statistical 
bootstrap of hadronic matter has 
lead from a different direction
to the consideration of the transition to
a hadron substructure phase \cite{Hag81,Hag84}.
The present day  lattice-QCD numerical 
simulations allows to evaluate  more rigorously 
the expected properties
of a equilibrium state of QGP, as we today 
call hot quark matter \cite{Kar00,Fod02a,Fod02b,Let03uj}.
QGP is the equilibrium state of matter at high temperature
and/or density. The question is if, in the short
time such conditions are available in laboratory 
experiments, this new state of matter can be created. There is
no valid first principles answer available today and 
we must address this issue experimentally. 

It is obviously very hard to probe experimentally the QGP phase which exists 
a very short time. It is not widely accepted  how  a reaction
involving formation of the deconfined state can be 
distinguished decisively from one involving  reactions  between 
individual confined hadrons. We observe in   the final state of a 
high energy interaction reaction,  irrespective if, or not,
deconfinement has been achieved,  a very large number of 
hadronic particles.  

The phenomenon of $J/\!\Psi$ and jet 
quenching   probes the early stage of the deconfined 
phase. Its relation to deconfinement
 has been  argued away in terms of effects of matter
density and changes in primary production reaction mechanisms. 
Plasma photon radiation is buried in the background of
final state photons from, {\it e.g.}, $\pi^0\to\gamma\gamma$. Dileptons
(that is in fact virtual photons) suffer from large experimental 
uncertainties, and a final state background which depends
of new physics as well (hadron properties at finite
temperature and density). 

Perhaps the most promising path to a convincing  evidence for
the formation of the deconfined state in relativistic
heavy ion collisions is the precision
study of  the  production  of soft hadrons. 
Insights into the nature of the hot hadronic matter fireball
are gained from  interpretation of an array of particle
 yields and spectra. A precise model description
allows to evaluate the global properties such as energy, 
entropy, strangeness
content of all particles produced. This opens up to further
investigation the properties of the matter  state 
that has hadronized.

Supporting this approach is the  study of 
several  reaction systems, {\it e.g.}, comparison of 
 $AA$ with $NN$ and/or $NA$ reactions.
In this case, the change of particle yield
when this contradicts the hadron rescattering model 
expectations is  evidence of new physics. Normally one
encounters in an  $AA$ experiment enhancement of expected yield 
comparing to $NN$ reactions. More recently, some have argued that 
the $AA$  yield of (strange) hadrons is `normal' and that the 
effect is due to a `suppressed'  $NN$ base yield. This argument does not
change the fact that in any case the  $AA$ yield does not arise 
from a  sequence of $NN$   reactions. 

\subsection{Survey of this work}\label{Survey}
The model we employ arises from the observation that hadron production 
is well described in a very large range of yields solely by evaluating
the phase space size.  Originally  proposed by 
Fermi in order to describe multiparticle production in high energy 
cosmic ray interactions \cite{Fer50,Fer53}, in 
the past 40 year the Fermi model
({\it i.e.}, statistical hadronization) 
has been tried successfully in many particle production environments,
beginning with $p$--$p$ collisions at various energies, and branching 
out to heavy ion as well as the elementary non-hadronic  $e^+e^-$
reactions.

We begin, in section \ref{StatHad},
 with a discussion of  the role  hadronic resonances play in  testing 
the principles of statistical hadronization.
We survey,  in subsection \ref{Res}, the  latest experimental 
hadronic resonance production
results from  RHIC. We next describe how particle
yields and ratios  can be obtained. We  introduce, in subsection \ref{fuga},
the ideas of hadrochemistry and show, in  subsection \ref{ssratios},
how particle ratios  determine hadronization parameters.
We turn to the study of phase space densities in
subsection \ref{anund}. We show that the entropy rich QGP phase 
hadronizes into  chemical non-equilibrium pion gas that can
absorb the entropy excess.
 For a more detailed discussion we refer to our
earlier reviews \cite{Let00b,Let02gp,Raf02jd}. 

We introduce  strangeness as signature of deconfined 
QGP phase in section \ref{SQGP}. We discuss the 
interest in strangeness both from experimental and theoretical
perspective in  subsection \ref{THR}. 
We introduce,  in subsection  \ref{2step}, the two step strange hadron 
production process, in which the  hadronization benefits from 
ample supply of strangeness already produced. This is  the origin of
multi strange hadron enhancement. We show, in subsection \ref{Mist},
 that the observed enhancement effect is   genuine. 
In subsection \ref{scool}, we discuss  how the 
outward  collective flow of matter influences the 
resulting fireball hadronization conditions and 
assists  the sudden hadronization.

Section \ref{hadSR} presents our findings about the
 hadronization  at SPS and at RHIC. We define first
the method of analysis in subsection \ref{Method} and follow this through with 
a discussion of RHIC fits in subsection \ref{RHICan}. 
We then address the energy dependence of the 
statistical parameters characterizing the fireball source 
in subsection \ref{StatPar}.  An interesting
result that follows is about the behavior of energy per baryon stopping.
Our primary interest is the strangeness yield which we 
address in  subsection \ref{StrangeS}.

Our study leads directly to the final question, is there a 
energy threshold for the onset of quark--gluon plasma formation,
which we address in  section~\ref{compare}.  
On important element of our study is to identify 
observables which are independent of the different baryon densities 
reached. We look from this perspective at the  kaon 
to pion ratio discontinuity reported at low end of SPS energy range. 
We show  that this is a result of baryon density effect in subsection \ref{Kpi}. 
In subsection \ref{LHC} we discuss a mechanism of kaon to pion
ratio enhancement arising from fast transverse expansion, and present 
the maximum enhancement expected at LHC. 
 A brief evaluation of our findings and their
importance follows in subsection \ref{final}.

\section{Statistical hadronization}\label{StatHad}
\subsection{Hadron resonances}\label{Res}
The initial  test  of  statistical hadronization 
approach to particle production is that 
within a particle `family',  particle yields with same valance quark
content are in relation to each other 
thermally  equilibrated. Thus the relative yield of, {\it e.g.},
$K^*(\bar s q)$ and $K(\bar s q)$ is  controlled only
by the particle masses $m_i$,  statistical weights (degeneracy) $g_i$ and the 
hadronization temperature $T$. In the Boltzmann limit
one has (star denotes the resonance): 
\begin{equation}\label{RRes}
{N^*\over N}= {g^*m^{*\,2}K_2(m^*/T)\over g\,m^{2}K_2(m/T)}.
\end{equation}
Validity of this relation implies insensitivity of the quantum matrix element 
governing the coalescence-fragmentation production of particles to
intrinsic structure (parity, spin, isospin), and  particle mass. 
The measurement of the relative yield of hadron resonances 
is a sensitive test  of the statistical hadronization hypothesis. 

The experimental ratio~\cite{Adl02pc,Fac02iz,Zha03}: 
$${(K^*+\overline{K^*})\over 2 K^-}=0.26\pm0.03\pm0.07,\qquad
{(K^*+\overline{K^*})\over 2 K^-}=0.20\pm0.01\pm0.03,$$
has been measured, left for most central collisions 
with $\sqrt{s_{NN}}=130$GeV, and right for most central 
200 GeV reactions, in RHIC-130, and RHIC-200 runs respectively
(65+65 GeV and 100+100 GeV per nucleon head on collisions 
of two nuclear  beams). The   RHIC  result naturally arises
from the ratio \req{RRes} allowing for resonance decays within
the statistical hadronization model 
for the temperature $T\simeq 145$ MeV. This is the favored
solution of a fit which allows for chemical 
non-equilibrium \cite{Raf02ga}, see subsection \ref{RHICan}. 

However, the  yield:  
$${\Lambda(1520)\over \Lambda}=0.022\pm0.01,$$
recorded in the RHIC-130 and RHIC-200 runs \cite{Mar02xi,Gau03},
is two times  smaller compared to  statistical
hadronization expectation. Statistical hadronization 
at higher temperature than $T=145$ MeV does not resolve this
discrepancy, but adds to it a 
$ K^*/K^-$ suppression.

These two ratios, seen together, are, at first, difficult to understand. 
Let us first note that an apparent  production suppression may 
actually be detectability  suppression. Namely, 
should  the decay products of resonances rescatter on 
other particles after formation, 
their energies and momenta will change. Hence,  not all produced 
resonances can be, in general, 
reconstructed from the decay products energies $E_i$ and
momenta $p_i$ by testing for the  invariant mass:
\begin{equation}\label{Resrec}
m^{*\,2}=(E_1+E_2)^2-(\vec p_1+\vec p_2)^2.
\end{equation}
The rescattering effect depletes more strongly the yields of 
shorter lived states. These decay sooner and are thus more within
the dense matter system. If resonance yield is as predicted
by statistical hadronization,  the decay life span
is expected to be larger then the hadronization time,  {\it e.g.}, 
$$
c\tau_{K^*}={\hbar c\over \Gamma_{K^*}}=3.9\mbox{\,fm\,}>c\tau_{\rm had}.
$$
This assures that most $K^*$ decay after thermal freeze-out of decay products 
($K,\pi$) in free space. On the other hand, we recall 
that the lifespan of $\Lambda(1520)$ is about 2.3 times longer 
than that of $K^*$. This means that contrary to the expectation derived from 
rescattering effect, it is  the longer lived $\Lambda(1520)$ 
state for which we find the signature to be 
 significantly depleted as compared to what could be maximally seen 
given  the statistical hadronization yield. 

A natural explanation of the $\Lambda(1520)$ signature suppression is 
in matter modification of the free space $\Gamma=15.6\pm1$ MeV
 decay width \cite{Raf01hp}. 
This width corresponds to $c\tau=12.6$\,fm. Thus for 
sudden hadronization, implied by the observed yield of  $K^*$,
$\Lambda(1520)$ decay should occur well outside 
the reaction domain.  However, if in medium width is at the 
level of $200$\,MeV, within a hadronization lifespan of 0.7fm/c 
half of the produced $\Lambda(1520)$ yield would become
unobservable due to rescattering of decay products. 
Such a resonance width of  $200$\,MeV is natural for a 400 MeV 
excited hadronic resonance state, once presence of the medium 
removes superselection constraints. 

Another  proposed  explanation of the $K^*, \Lambda(1520)$ riddle 
is that there is $K^*$ yield enhancement by regeneration \cite{Ble02dm}. 
This $\pi+K\to K^*$ production  mechanism is 
generally associated with a long lived hadron phase. 
Factor 5 enhancement over statistical 
hadronization yield is needed so that given the life span
difference,  the apparent reduction  of  $\Lambda(1520)$ by factor 2
is consistent with the observed high $K^*$ yield. 
We are  assuming here that the regeneration of $\Lambda(1520)$  
by $N,K$ collisions is not significant. However,
a large $K^*$ yield enhancement contradicts 
 basic   principles,  in that the heavy  $K^*$ would    
compete with $K$ in the overall bare yield. For this reason,
we do not believe that this is a viable explanation.

We conclude that the statistical hadronization is doing reasonably 
well, the predictions in general agree with experiment in considerable 
detail, and where strong deviation is observed ($\Lambda(1520)$) 
a good reason for this is at hand. 
Resonances test  both statistical hadronization, and  the
nature of the hadronization process \cite{Tor2001tg}, 
 and open a path to study in medium modifications of particle 
hadronic widths.

\subsection{Chemical fugacities}\label{fuga}
The freeze out temperature determines the shape of initially 
produced particle spectra, and thus 
it has considerable influence over their yield as well. However,
the normalization of the spectra and thus particle yields are more 
directly  controlled by the particle fugacity 
$\Upsilon_i\equiv e^{ \sigma_i  /T}$, where $ \sigma_i$
 is particle `$i$' chemical potential. Since for each related  
particle and antiparticle  pair we need two chemical potentials, it 
has become convenient to choose parameters such that we can control 
the difference, and sum of these separately. For example for nucleons
and antinucleons $N,\overline{N}$ the
two chemical factors are  chosen as: 
\begin{equation}
\sigma_{N}\equiv \mu_b +T\ln\gamma_N ,\qquad
\sigma_{\overline{N}}\equiv -\mu_b +T\ln\gamma_N,
\end{equation}
\begin{equation}
\Upsilon_N=\gamma_N e^{ \mu_b /T}, \qquad\qquad
\Upsilon_{\overline{N}}=\gamma_N  e^{- \mu_b /T}.
\end{equation}

The role of the two factors can be understood
considering the first  law of thermodynamics: 
\begin{eqnarray}
dE+P\,dV-T\,dS&=&\sigma_N\,dN+
      \sigma_{\overline{N}}\,d\overline{N},\\ \nonumber
&=&
 \mu_b (dN-d\overline{N})+ T\ln \gamma_N (dN+d\overline{N}).
\end{eqnarray}
The   (baryo)chemical potential 
 $\mu_b$,  controls the baryon number, arising from the particle difference. 
$\gamma_N$, the phase space occupancy,   regulates the number of nucleon--antinucleon pairs present. 

There are many different hadrons, and in principle, we should
assign to each a chemical potential and then look for 
chemical reactions which relate these chemical potentials. 
However, more direct way to accomplish
the same objective consists in  characterizing 
each particle by the valance quark content \cite{Koc83}, 
forming a product of chemical factors,  {\it e.g.},  for  $p(uud)$,
\[
\Upsilon_{p(uud)}=\gamma_u^2\gamma_d\ \lambda_{u}^2\lambda_{d},\qquad\qquad
\Upsilon_{\bar p(\bar u\bar u\bar d)}=\gamma_u^2\gamma_d\ \lambda_{u}^{-2}\lambda_{d}^{-1}.
\] 

Considering the isospin symmetry of strong interactions 
aside of the three quark fugacities  $\gamma_i, i=u,d,s$, we also
introduce the light quark  fugacity:
\beql{lamq}
\lambda_q^2=\lambda_u\lambda_d,
\qquad
\lambda_b=\lambda_q^3.
\eeq
To transcribe the fugacities into chemical potentials we recall:  
\beql{muq}
\lambda_{i}=e^{\mu_{i}/T},\qquad \mu_q=\frac12 (\mu_u+\mu_d).
\eeq
The relation between quark based chemical potentials and the
two principal  hadron based chemical potentials of baryon number and 
hadron strangeness $\mu_i, i=b,{\rm S}$ is:
\beql{muhadq}
\mu_b=3\mu_q\qquad
\mu_s=\frac 1 3 \mu_b-\mu_{\rm S},
\qquad \lambda_s={\lambda_q\over\lambda_{\rm S}}.
\eeq

An important (historical)  anomaly is the negative 
S-strangeness in $s$-hadrons, {\it e.g.}:
\[
\Upsilon_\Lambda=\gamma_u\gamma_d\gamma_s\,e^{(\mu_u+\mu_d+\mu_s)/T}
           =\gamma_u\gamma_d\gamma_s e^{(\mu_b-\mu_{\rm S})/T},
\]
\[
\Upsilon_{\overline\Lambda}=\gamma_u\gamma_d\gamma_s\,e^{(-\mu_u-\mu_d-\mu_s)/T}
   =\gamma_u\gamma_d\gamma_s\,e^{(-\mu_b+\mu_{\rm S})/T}.
\]

There are  two types
 of chemical factors $\gamma_i$ and $\mu_i$ and thus
two types of chemical equilibria. These are shown 
 in table~\ref{parameters}. The absolute
equilibrium is reached when the phase space occupancy approaches unity,
$\gamma_i\to 1$. The distribution of flavor (strangeness) among many
hadrons is governed by the relative chemical equilibrium. 

\begin{table}[t]
\caption{\label{parameters}Four quarks $s,\ \overline{s},\ q,\ \overline{q} $
 require four chemical parameters; right: name of the associated chemical equilibrium} 
\vskip 0.3cm
\begin{center}
\begin{tabular}{ll|l}
\hline
\hline 
 $\gamma_{i}$&   controls overall abundance & Absolute   chemical\\
&of quark  ($i=q,s$)  pairs & equilibrium\\
\hline
 $\lambda_{i}$&   controls difference between & Relative   chemical\\
&quarks and antiquarks  ($i=q,s$) & equilibrium
\end{tabular}
\end{center}
\end{table}

There is considerable difference in the dynamics of  these two particle 
yield equilibration. This can be, {\it e.g.},  understood 
considering  strangeness in the hadronic gas phase.  
The two principal chemical processes are seen in \rf{exchange}.
The  redistribution of strangeness among 
(in this example) $\Lambda,\,\pi$ and $N,$ K  seen on left in \rf{exchange}.
constitutes approach to the relative  chemical equilibrium 
of these species. The $s,\bar s$ pair
 production process, on right in \rf{exchange}, 
is responsible for absolute   
chemical equilibrium of strangeness. Achievement of the 
absolute equilibrium, $\gamma\to 1$, require  more rarely 
occurring truly inelastic collisions with creation of new particle
pairs. 

\begin{figure}[h]
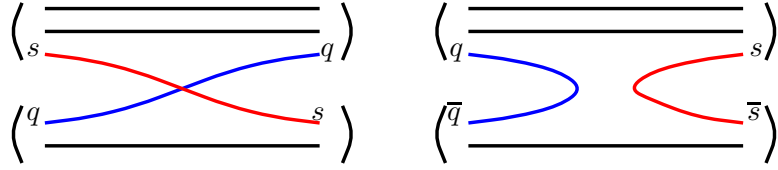
 
\begin{center}
\psfig{width=4.5cm,figure=\pathnow exchange.eps}
\hspace*{1cm}\psfig{width=4.5cm,figure=\pathnow produc.eps}
\end{center}
\vspace{-2.3cm}
\hspace*{1.4cm}${ s}$ \hspace*{3.6cm}${ q}$
\hspace*{1.4cm}${q}$\hspace*{3.8cm}${ s}$\\[0.4cm]
\hspace*{1.4cm}${ q}$\hspace*{3.6cm}${ s}$
\hspace*{1.5cm}${ \overline{ q}}$\hspace*{3.8cm}${ \overline{ s}}$
\vspace{0.5cm}
\caption{\label{exchange}
Typical strangeness exchange (left) and production (right)
reactions in the hadronic gas phase.}
\end{figure}

\subsection{Ratios of particle yields}\label{ssratios}
When particles are produced in hadronization, 
we speak of chemical freeze-out. 
To obtain the particle yields up to overall normalization 
the individual Fermi--Dirac and Bose--Einstein phase
space integrals are evaluated. In order to arrive at the full yield, 
one has to be sure to include all the hadronic resonance
decays feeding into the  yield considered, {\it e.g.}, the decay 
 $K^*\to K+\pi$ feeds into $K$ and $\pi$ yields. 
This actually constitutes a book keeping challenge in study of particle 
multiplicities, since decays are contributing at the 50\% level to 
practically all particle yields. Sometimes the decay contribution
can be dominant, as is generally the case for the  pion yield. 
For pions in particular, 
each resonance contributes relatively little in the final count, it is
the large number of resonances that contribute which  competes with the 
direct pion yield. 

To determine the parameters governing the chemical 
freeze-out, we analyze particle yields in terms of the chemical 
parameters and the temperature. 
Except for direct pions practically always one can use Boltzmann 
approximation and large reaction  volume, and what follows 
in this subsection assumes that this simple situation applies. 

It is often appropriate to study ratios of particle yields as these
can be chosen such that certain physical features can be isolated. For
example, just the two ratios,  
\beql{bLL}
R_\Lambda = 
\frac{\overline{\Lambda}+\overline{\Sigma}^0+\overline{\Sigma}^*+\cdots }{
{\Lambda}+{\Sigma}^0+{\Sigma}^*+\cdots}=\frac{\bar s\bar q\bar q}{sqq}=
\lambda_{\rm s}^{-2} \lambda_{\rm q}^{-4} =
e^{2\mu_{\rm S}/T}e^{-2\mu_b/T}\, ,
\eeq
\beql{bXX}
R_\Xi =  
\frac{\overline{\Xi^-}+\overline{\Xi^*}+\cdots }{{\Xi^-}+{\Xi^*}+\cdots}=
\frac{\bar s\bar s\bar q}{ssq}=
 \lambda_{\rm s}^{-4} \lambda_{\rm q}^{-2}=e^{4\mu_{\rm S}/T}e^{-2\mu_b/T} \, ,
\eeq 
lead to a very good estimate of the baryochemical potential and strange
chemical potential~\cite{Raf91a}, and thus to predictions of other particle 
ratios.

The   sensitivity to phase space occupancy 
factors $\gamma_i$  derives from comparison of hadron yields
with differing $q,s$ quark content, {\it e.g.}:
\beql{XiLam}
\frac{\Xi^-(dss)}{\Lambda(dds)}\propto 
\frac{\gamma_d\gamma_s^2}{\gamma_d^2\gamma_s}\
\frac{g_\Xi\lambda_d\lambda_s^2}{g_\Lambda\lambda_d^2\lambda_s}\,,\qquad
\frac{\overline\Xi^-(\bar d\bar s\bar s)}{\overline\Lambda(\bar d\bar d\bar s)}
\propto 
\frac{\gamma_d\gamma_s^2}{\gamma_d^2\gamma_s}\
\frac{g_\Xi\lambda_d^{-1}\lambda_s^{-2}}{g_\Lambda\lambda_d^{-2}\lambda_s^{-1}}\,.
\eeq

In \req{XiLam}, each of the ratios also contain chemical
 potential factors $\lambda_i$. These can be eliminated 
by taking the product of particle
ratio with antiparticle ratio, thus 
\beql{Xi2Lam}
\frac{\Xi^-(dss)}{\Lambda(uds)}
\frac{\overline\Xi^-(\bar u\bar s\bar s)}{\overline\Lambda(\bar d\bar d\bar s)}
=C_{\Xi\Lambda}^2
\left(\frac{\gamma_s}{\gamma_u}\frac{g_\Xi}{g_\Lambda}\right)^2\,.
\eeq
The proportionality constant $C_{\Xi\Lambda}^2$ 
describes the phase space size ratio for the two 
particles of different mass. It incorporates the contributions
from resonance decays, which of course differ from particle to particle. 

The method applied in \req{Xi2Lam} can be used in several other such double 
particle ratios. The relevance of this is that we have identified an experimental 
observable (combination of particle ratios) 
solely dependent on two parameters of statistical hadronization
and chemical freeze-out, the temperature $T$ which controls the phase space
factor ratio $C$ and the ratio $\gamma_s/\gamma_q$.
Another double ratio of considerable importance is
\beql{Lam2p}
\frac{\Lambda(uds)}{p(uud)}
\frac{\overline\Lambda(\bar u\bar d\bar s)}{ \bar p\,(\bar u\bar u\bar d)}
=C_{\Lambda p}^2
\left(\frac{\gamma_s}{\gamma_u}\frac{g_\Lambda}{g_p}\right)^2\,.
\eeq
Both  ratios 
Eqs.\,(\ref{Xi2Lam},\ref{Lam2p}) are  independent of 
chemical potentials, their measurement  allows to constrain the 
value of $\gamma_s/\gamma_q$ only as function of $T$ \cite{Raf02ga}.

Double  ratio made of mesons in general will be weakly dependent on chemical 
potentials since some  kaons and pions 
 are decay products of baryonic  resonances. The leading term is in general 
originating in  direct and resonance mesons, for example:  
\beql{K2pi}
\frac{K^+(u\bar s)}{\pi^+(u\bar d)}
\frac{K^-(\bar u  s)}{p^-(\bar u  d)}
=C_{K\pi}^2
\left(\frac{\gamma_s}{\gamma_q}\frac{g_K}{g_\pi}\right)^2+\cdots \,.
\eeq

In summary, and  for the benefit of our later  discussion in subsection \ref{Kpi}, 
we remember that  there is no dependence in Eqs.\,(\ref{Xi2Lam},\ref{Lam2p})
and almost none in Eq.\,(\ref{K2pi}) on the 
baryochemical potential and strange chemical potential. 
This eliminates dependence of the baryon and strangeness density, and allows
to focus on the physics issues unrelated to baryon compression effects.

\subsection{Particle phase space density and hadronization}\label{anund}
The maximization of microcanonical  entropy, 
\beql{entrof}
{S}_{F,B}=
 \int\!\frac{d^3\!p\, d^3\!x}{(2\pi\hbar)^3}\,
   \mp\left[(1\mp f_{F,B})\ln(1\mp f_{F,B})-   f_{F,B}\ln f_{F,B}\right]\,,
\eeq
(where indeed minus sign is for fermions and plus sign for bosons) 
subject to energy and baryon number conservation 
implies the quantum distributions~\cite{Let93qa}:
\beql{BosFer}
 {{d^6N_i}\over{d^3pd^3x}}= {g_i\over (2\pi)^3}
{1\over  \Upsilon_i^{-1} e^{E_i/T}\pm 1},\quad
 \Upsilon_i^{\rm bosons} \le e^{m_i/T}.
\eeq
When the phase space is not densely occupied, $\pm 1$ in the denominator 
can be neglected and we have the Boltzmann approximation we used in last subsection:
\beql{Bol}
 {{d^6N_i}\over{d^3pd^3x}}=g_i{ \Upsilon_i  
\over (2\pi)^3}e^{-E_i/T}.
\eeq

In \req{BosFer}, for the Boson distribution, the value of $\gamma_q$ is
limited by the condensation singularity.
The  maximum value of $\gamma_q^{\rm max}=e^{m_\pi/(2T)}$ plays a very pivotal role
considering that the mass of the pion and hadronization temperature 
are similar. 
Large value of $\gamma_q\to e^{m_\pi/T}$ can be directly 
noticed in pion spectra in an uptilt in the soft portion
of the $m_\bot$ distribution. 
A similar constraint is also  arising from kaon condensation for $\gamma_s$
but it is much  less restrictive:
\begin{equation}\label{Kcond}
{\gamma_s\over\gamma_q}\left(\lambda_s\over\lambda_q\right)^{\!\!\pm 1}
<e^{(m_{\rm K}-m_\pi)/ T}\simeq 11\,.
\end{equation}

In the local restframe the  particle yields  are 
proportional to the momentum integrals of the distribution \req{BosFer}. 
As example, for  pions $\pi,$ nucleons $ N$ and antinucleons 
$\overline N$ we have:
\begin{eqnarray}\label{Npi}
N_\pi &=& C{V} g_\pi\!\!\int_R\frac{d^3p}{(2\pi)^3}
  \frac{1}{\gamma_q^{ -2}e^{\sqrt{m_\pi^2+p^2}/T}-1}\,,
     \qquad \gamma_q^2<e^{m_\pi/T},\\[0.3cm]
N&=&C {V}g_N\!\!\int_R\frac{d^3p}{(2\pi)^3}
   \frac{1}{1+\gamma_q^{ -3}\lambda_q^{ -3 }e^{E/T}},\\[0.3cm]
\overline {N}&=&  C{V}g_N\!\!\int_R\frac{d^3p}{(2\pi)^3}
    \frac{1}{1+\gamma_q^{ -3 }\lambda_q^{ +3 }e^{E/T}}.
\end{eqnarray}
Naively,
we would think that the coefficient is simply the volume $V$, as it
would be the case for a gas of hadrons. However, more generally there
is a common additional factor $C$ which is determined by the dynamics of 
the hadronization process. The region of integration $R$ is determined in
terms of the experimental acceptance, keeping in mind that we
are here considering the phase space in rest frame of hadronizing QGP,
while the experimental detector is at rest in the laboratory. Both
the motion of the collision center of momentum frame with 
respect to the detector, and the collective flow of QGP have to 
be allowed for. 

Generally a small region of particle rapidity $y$ but 
nearly all of range of $m_\bot$ is accepted in experiments
at RHIC. Assuming the yield  of particles is practically constant 
as function of $y$, it is possible to imagine that the 
yield arise from a series of fireballs placed at 
different rapidities. Thus in this 
limit we can proceed to study hadronization as if we had
a full phase space coverage. However, in this case in particular
the proportionality constant  $C$ can be quite different from unity.

The hadronization of  QGP phase
 is not fully understood. However, numerous experimental
indicators show that hadrons are emerging rapidly from a relatively 
small space--time volume domain. In such a case, hadron formation has 
to absorb the high
entropy  content of QGP which originates in broken color bonds.
The lightest hadron is pion and most entropy per energy 
is consumed in hadronization  by producing these
particles abundantly. We evaluate \req{Npi} to find particle
number, and  the entropy content follows from \req{entrof}:
\begin{equation}
{S}_\pi=
 \int\!\frac{d^3\!p\, d^3\!x}{(2\pi\hbar)^3}\,
   \left[(1+f_\pi)\ln(1+f_\pi)-   f_\pi\ln f_\pi\right]\,,
\end{equation}
\begin{equation}
f_\pi(E)=\frac{1}{\gamma_{q}^{-2}e^{E_\pi/T}-1}
\,,\quad 
E_\pi=\sqrt{m_\pi^2+p^2}.
\end{equation}

As is seen in \rf{JRABSSNE}, the maximum  entropy 
density $S/V$ occurs for  an oversaturated  pion gas, 
$\gamma_q\simeq e^{m_\pi/2T}\simeq 1.6$. Here, 
the entropy density of such a saturated Bose  gas
is twice as large  as that of
chemically equilibrated Bose gas. Since aside of pions also
many other hadrons are produced, the large value of $\gamma_q$ 
is necessary and sufficient to allow for the smooth in $\mu_b, T, V$ 
transformation  of a QGP into hadrons. The
number of active degrees of freedom in the oversaturated hadron gas
with $\gamma_q\to \gamma_s^{\rm max}$ and in `freezing' QGP phase is
very similar.

\begin{figure}[h]
\vspace*{1.9cm}
\hspace*{0.7cm}
\epsfig{width=10cm,height=7cm,figure=\pathnow 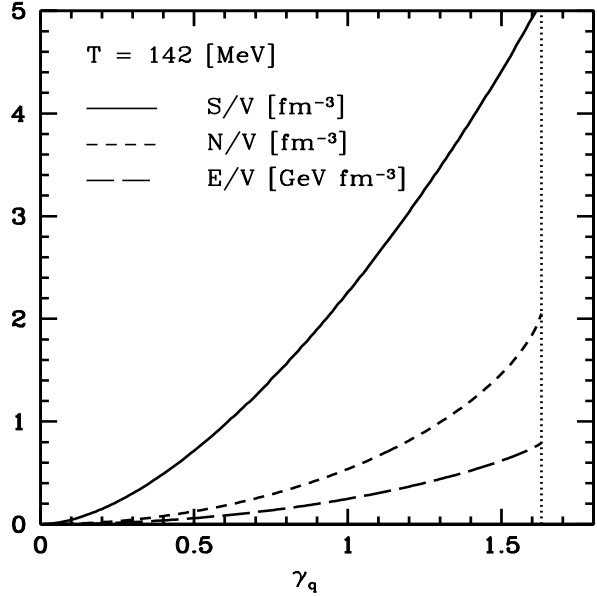}
\vskip -.8cm
\caption{\label{JRABSSNE}
Entropy density $S/V$ along with particle density $N/V$ and energy density 
$E/V$ as function of $\gamma_{q}$ at $T=142$\,MeV. 
}
\end{figure}

\section{Strangeness --- a popular QGP diagnostic tool} \label{SQGP}
\subsection{Motivation for study of strangeness}\label{THR}
It is widely agreed that the process of strange particle production
in relativistic nuclear collisions offers considerable insight into the 
structure and dynamics of the dense matter formed in these reactions.
The experimental reasons for the interest are practical, as this 
observable is experimentally accessible. The 
theoretical interest has been motivated by the recognition that
strangeness is a marker of the thermal gluonic degree of freedom
of the deconfined phase \cite{RM82,Raf82}. These result
is enhanced abundance of strangeness, and thus production of 
numerous strange hadrons \cite{abundance}.

In an nutshell the experimental motivation for study of strangeness is:
\begin{itemize}
\item  There are many strange particles allowing to study 
 different physics questions $(q = u,\ d)$:
\[\!\!\!\!
\phi(s\bar s),\  K(q\bar s),\  \overline{K}(\bar q s),
\  \Lambda(qqs),\  \overline{\Lambda}(\bar q\bar q\bar s),
\  \Xi(qss), \  \overline{\Xi}(\bar q\bar s\bar s),
\  \Omega(sss), \  \overline{\Omega}(\bar s\bar s\bar s);
\]
\item Strange hadrons are subject to a self analyzing decay
within a  few cm from the point of production. As an example, we see
the cascading decay of a doubly strange $\Xi^-$ in \rf{VSIG};
\begin{figure}[h]
\hspace*{2.5cm}
\psfig{width=8cm,figure=\pathnow 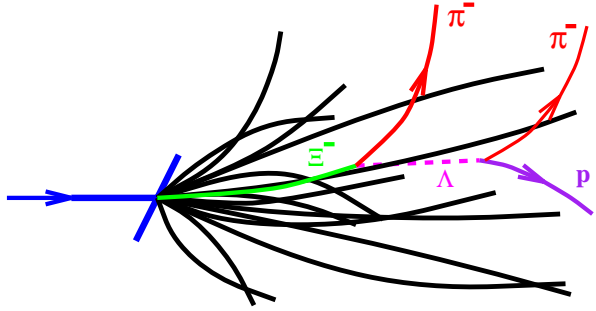}\vskip 0.cm
\caption{\large\baselineskip 0.5cm
$\Xi^-$-decay, dashed line the invisible $\Lambda$ emerging 
from the decay kink ending in the decay `V' of the final state charged 
particles. \label{VSIG}
}
\end{figure}

\item Though some of the strange hadrons are produced quite 
rarely in $p$--$p$ reactions, the enhancement we encounter in $A$--$A$
reactions produces `exotic' strange observables with yields sufficient to 
reach  a relatively high  statistical  significance.
\end{itemize}

The theoretical case for study of strangeness  has been researched in depth:
\begin{itemize}
\item Production of strangeness occurs predominantly 
in gluon fusion $gg\to s\bar s$. Thus abundant
strangeness is linked to presence of thermal gluons from QGP. The
quark process \cite{Bir82},  $q\bar q\to s\bar s$, contributes at level of 
10--15\% of total rate, the Feynman diagrams for the lowest order
process are seen in \rf{SSPROD};
\begin{figure}[t]
\centerline{\hspace*{1cm}\psfig{height=8cm,angle=-90,
figure=\pathnow 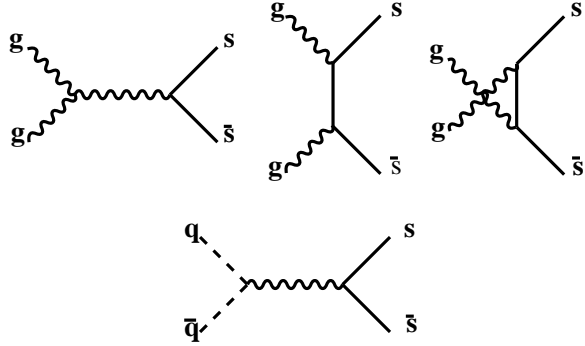}}
\caption{ \label{SSPROD} Lowest order Feynman diagrams for
the production of flavor. Top line: gluon fusion and  bottom
line:  quark--antiquark fusion into strange quark pair. 
}
\end{figure}

\item Coincidence of scales: $m_s\simeq T_c$ implies that the 
relaxation time required to equilibrate strangeness yield 
chemically, $\tau_s$, is of the same magnitude as the 
lifespan of QGP,  $\tau_s \simeq \tau_{\rm QGP}$. In consequence, 
strangeness acts as   a clock for the precise QGP lifespan.
\item Due to presence of baryon density, the yield of light antiquarks is
suppressed compared to yield of quarks and one thus can encounter
$\bar s>\bar q$. Ensuing recombination production of antibaryons 
at hadronization of QGP can lead to a relative 
 strange  antibaryon enhancement, and at RHIC to
 (anti)hyperon dominance of (anti)baryons \cite{Raf99a}.
\end{itemize}

\subsection{Two step hadron formation mechanism in QGP}\label{2step}
The establishment of a ready supply of strange quarks 
occurs in a manner independent of the production of final
state hadrons. After some time has passed, QGP drop 
reaches hadronization condition in its internal pressure driven 
expansion. Then, the final state
hadrons  emerge, as is seen in \rf{JRSTRANGEPROD}. Thus
 strange hadron production occurs in two independent 
steps following each other in time:
\begin{enumerate}
\item $gg\to s\bar s$ predominantly  in the early hot QGP;
\item hadronization of pre-formed  $s,\,\bar s$ quarks after 
QGP cools.

\end{enumerate}
\begin{figure}[h]
\vskip -0.5cm
\hspace*{2cm}\epsfig{width=9cm, angle=-90,figure=\pathnow 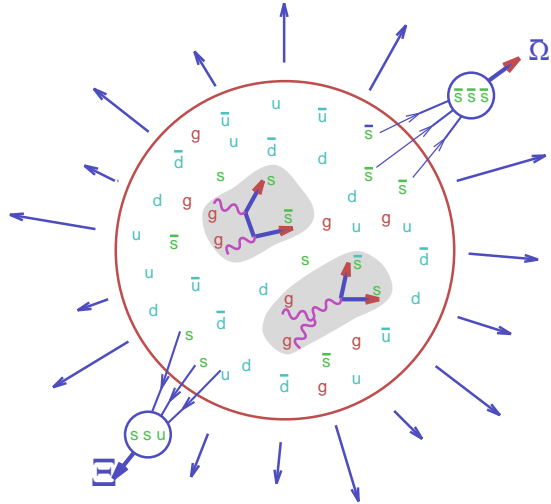}
\vskip -1.5cm
\caption{Illustration of the cross-talk two step mechanism of strange hadron formation 
from QGP: inserts show gluon fusion into
 strangeness, followed by QGP recombinant hadronization.
\label{JRSTRANGEPROD}}
\end{figure}

This sequence allows the production  of complex rarely produced 
multi strange (anti)particles, enabled by 
 `cross talk' between   $s,\,\bar s$ quarks made
in different microscopic reactions. We consider observation
of this process to be a specific signature of 
deconfinement, since it provides evidence for  both
the activity of thermal gluons, and the mobility of strange quarks.
The specific feature of such an enhancement is that the 
enhanced  production of strange
antibaryons is increasing with strangeness content.

\subsection{Strange hadron enhancement or suppression?}\label{Mist}
The enhancement of strange hadrons has been  studied as function of 
the number of participating baryons, which are evaluated
in terms of `wounded nucleons', that is nucleons which 
have been involved in an inelastic scattering in the 
collision. We show, in \rf{NA57res}, the 
results of NA57 \cite{Ant02,Man03}, which extend those presented 
by WA97 collaboration.
\begin{figure}[tbh]
\psfig{height=7.6cm,clip=1,figure=\pathnow 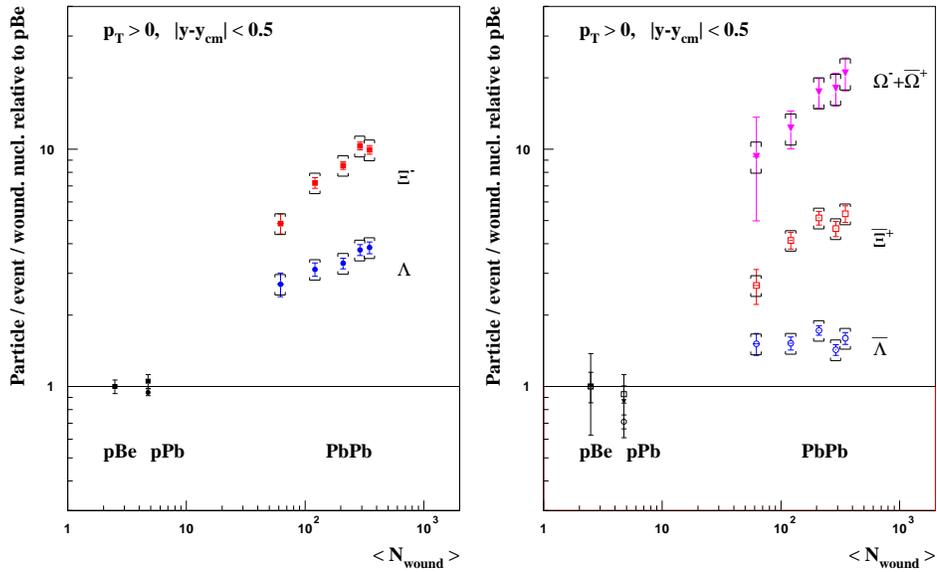}
\vspace*{-0.3cm}
\caption{
\label{NA57res}
Specific yield per wounded nucleon
of strange baryons $\Lambda, \Xi$ and antibaryons 
 $\overline\Lambda, \overline\Xi$ and of $\Omega+\overline\Omega$
as function  of the number of wounded nucleons \protect\cite{Man03}. 
 The data are
obtained in 5 centrality bins which cover the range 50--350
participants. The baseline for the enhancement is the yield 
observed in the $p$--Be reactions.
}
\end{figure}
The yield of strange hadrons shows the expected increase of
enhancement  as function of newly made quark content $s,\bar s,\bar q$,
and as function of the participant number (see for example the result
shown in figure 37, in \cite{Raf96}.  The enhancement is in  \rf{NA57res}
reported with a base for  $p$--Be reactions. 
It has been  recognized  long ago that 
the relatively small number of strange quarks produced
 in  proton induced reaction could require evaluation of statistical  yields
applying microscopic particle number constraints
 of  relevance in such a limit \cite{Raf80}. 
Thus  statistical equilibrium yields for the 
 $p$--$p$ reaction systems are suppressed  and
thus enhancement is seen comparing $AA$ to $pp$. However, 
results of NA57 we see in  \rf{NA57res} disagree with the
expected behavior of this suppression \cite{Ham00c}.

 Without going into 
the theoretical details discussed elsewhere \cite{Raf01bu},
the important feature of the micro-canonical yield suppression 
is that  the asymptotic particle
yields are relatively quickly attained as is seen in \rf{CanEnh}. 
We see here the enhancement $E_i$  generated by rebasing the small
system yields to unity for the strangeness content $i=1,2,3$. 
The top of the figure shows the resulting enhancement for the 
 $p$--$p$  reaction system while the bottom estimates the result
for  $p$--Be as used by NA57, assuming that the yield of strange
quarks pairs $\langle N_s\rangle$ doubles  from 0.66 to 1.3.
The size dependence is expressed in terms of the canonical yield
of strange quarks produced (thus the base enhancement 
unity   is at the value $N^{\rm CE}=0.66$
in the top portion of \rf{CanEnh}, while it is at $N^{\rm CE}=1.3$
 in the bottom portion. 

\begin{figure}[tbh]
\vspace*{-0.2cm}
\hspace*{2.5cm}
\psfig{height=8.5cm,clip=1,figure=\pathnow 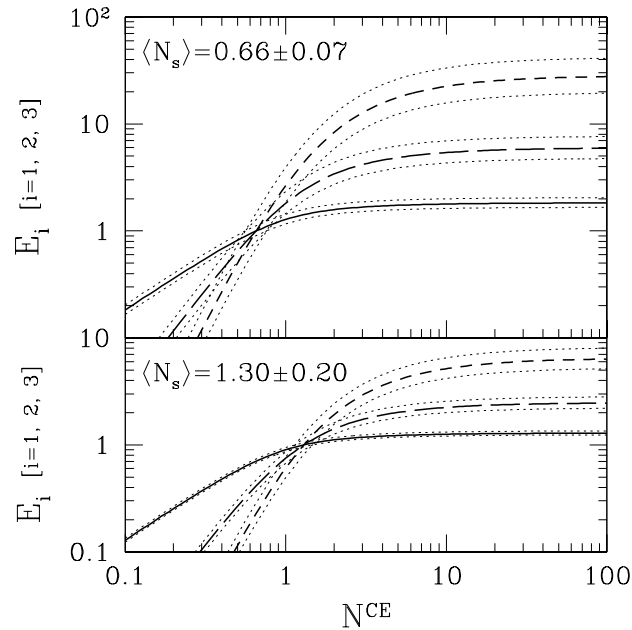}
\caption{\label{CanEnh}
Enhancement $E_i$ of strangeness $i=1$ (solid lines),
$i=2$ (long dashed lines) and $i=3$ (short dashed lines) 
hadrons   as function
of the canonical yield of strangeness $N^{CE}$. The cases 
for  $p$--$p$ (top panel) and $p$--Be  (bottom panel) 
reactions are shown. 
}
\end{figure}

Looking at the results in \rf{NA57res}, 
we see that systems with 100--300 baryons are still growing in 
yield. On the other hand we see in \rf{CanEnh} that the
 saturated yield of strange hadrons is attained  at 
$N^{CE}=5$, which yield is attained  for a fireball with 
10--30 baryons. This, along with the other issues \cite{Raf01bu},
raised earlier shows that the enhancement effect is indeed
not due to a suppression of the base of comparison, but 
is due to kinetic processes already described, {\it i.e.}, excess
strangeness production in QGP and subsequent 
 hadronization process. 

\subsection{Sudden hadronization}\label{scool}
A fireball is  not a piece of deconfined matter 
sitting still. Given its origin in a collision on two
atomic nuclei, it is initially  made of 
highly compressed matter. Thus it subsequently undergoes a
rapid collective explosive outflow. Such a collective motion of 
color charged quarks and gluons contributes an important 
collective component in the pressure, beyond the rest frame 
thermal pressure. This can be seen considering the stress portion of the 
energy-momentum tensor: 
\begin{equation}
T^{ij}=P\delta_{ij}+(P+\varepsilon)\frac{v_iv_j}{1-\vec v^{\,2}}\,.
\end{equation}
The rate of  momentum flow vector  $\vec {\cal P} $ 
at the surface of the fireball
 is obtained from the energy-stress tensor  $T_{kl}$: 
\begin{equation}
\vec {\cal P}\equiv \widehat{\cal T}\cdot \vec n=P \vec n+(P+\varepsilon)
  \frac{\vec v_{\mbox{\scriptsize c}}\, \vec v_{\mbox{\scriptsize c}}\!\cdot\! \vec n}
          {1-\vec v_{\mbox{\scriptsize c}}^{\,2}}\,.
\end{equation}
The pressure  and energy  comprise both the particle and the  vacuum properties:
 $$P=P_{\mbox{\scriptsize p}}-{\cal B}\qquad 
\varepsilon  =\varepsilon_{\mbox{\scriptsize p}}+{\cal B}\,.$$

The condition  $\vec {\cal P}=0$  reads:
\begin{equation}
{\cal B}\vec n=P_{\mbox{\scriptsize p}}\vec n+
      (P_{\mbox{\scriptsize p}}+\varepsilon_{\mbox{\scriptsize p}})
\frac{\vec v_{\mbox{\scriptsize c}} \, \vec v_{\mbox{\scriptsize c}}\!\cdot \!\vec n}
        {1-v_{\mbox{\scriptsize c}}^{2}}\,.
\end{equation}
Multiplying with  $\vec n$ , we find:
\begin{equation}\label{Kappa}
{\cal B}=P_{\mbox{\scriptsize p}}+
      (P_{\mbox{\scriptsize p}}+\varepsilon_{\mbox{\scriptsize p}})
\frac{\kappa  v_{\mbox{\scriptsize c}}^2}{1-v_{\mbox{\scriptsize c}}^{2}}\,,
\qquad
\kappa=\frac{(\vec v_{\mbox{\scriptsize c}}\cdot 
             \vec n)^2}{v_{\mbox{\scriptsize c}}^2}\,.
\end{equation}
We note that the hadronization hypersurface, and in particular
the  angular relation between the collective flow direction 
vector $v_{\mbox{\scriptsize c}}$  and surface normal direction,
defines the force balance condition seen in \req{Kappa}. 

In order 
to satisfy the condition \req{Kappa}, we must have 
$P_{\mbox{\scriptsize p}}<{\cal B}$: the  QGP phase pressure 
$P=P_{\mbox{\scriptsize p}}-{\cal B}$ must 
be {\em negative} at hadronization since it has to compensate the
effect of the positive dynamical pressure. When $P\to 0$ 
and a nonzero velocity of collective flow remains pointing  outward, 
the surface is torn apart in a rapid filamentation instability. 
This situation can {\em only} arise since the quark--gluon 
matter presses again the collective vacuum which is not subject  
to collective dynamics \cite{Raf00}. In this aspect the dynamics
of QGP expansion reminds us of a gas bubble growth within 
a liquid (cavitation phenomenon). 

The hypothesis  that hadronization of the QGP deconfined phase formed
in high energy nuclear collision is sudden has
a relatively long history.  In the 1986 review Koch {\it et al.} \cite{Koc86ud} note
that strange hadrons will be viable observables of QGP should a long
lived  hadron reequilibration not occur.  They offered a study of  hadron 
 abundances arising from sudden hadronization  of strangeness rich QGP. 
Once data became available, the first analysis of the experimental  
hadron abundances 
is carried out within this approach \cite{Raf91a}. The important feature of sudden
hadronization, the identity of temperatures of diverse hadrons and 
in particular of baryons and antibaryons is recognized. Shortly
after dynamical  models of sudden hadronization begin to emerge 
\cite{Cso94,Cse95}.

Phase boundary for a system at rest
between the hadron gas domain and quark--gluon plasma in the
$\mu_b,T$ plane, is shown in \rf{PLTMUBLIQ3}. The solid thin line
is that for hypothetical point hadrons, while dashed line shows 
the typical estimate for  finite volume hadrons. The effect of the 
  `wind' of flow of QCD matter
is also  seen  in \rf{PLTMUBLIQ3}, for a geometry of 
hadronization  with  $\kappa=0.6$.

\begin{figure}[htb]
\vspace*{-1.5cm}
\hspace*{2cm}\psfig{width=9.cm,clip=,figure=\pathnow 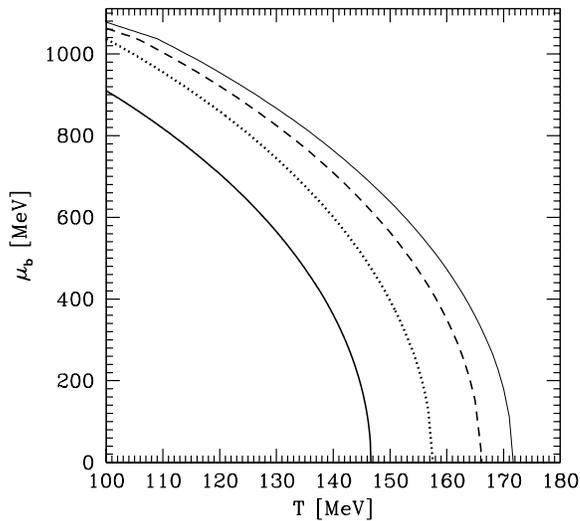}
\vspace*{-0.8cm}
\caption{\label{PLTMUBLIQ3}
Hadronization boundary in the
$\mu_b,T$ plane, QGP-HG transition for  point hadrons (solid thin line)
and finite volume hadrons (dashed line), 
are typical estimates of QGP--HG boundary  for a system at rest. 
Dotted line for finite size hadrons includes the effect of the 
collective flow velocity for $v_c=0.32$, $\kappa=0.6$.
 Thick solid line: breakup with 
$v=0.54$, $\kappa=0.6$
(based on \protect\cite{Raf00}).  
}
\end{figure}

As the expansion velocity increases, the magnitude of the color 
wind effect increases with velocity and is here shown for  
$v_c=0.32$ (dotted)   and $v_c=0.54$ (solid lines).
With increasing flow velocity the phase boundary 
 moves to lower temperatures. This effect reduces
for a given $\mu_b$ the magnitude of critical temperature in a 
significant way.  

As collective flow velocity decreases, the hadronization temperature 
increases. Even though this effect is somewhat compensated by the 
influence of the  increase in $\mu_b$ which implies a decrease 
in $T$, this  `normal' curvature effect  seen 
in each of the lines in  \rf{PLTMUBLIQ3} is smaller than the 
`new' flow effect as we move from line to line in  \rf{PLTMUBLIQ3}. 
At RHIC,  the largest transverse velocity is reached, and thus 
largest supercooling achieved. We expect the largest effect from
the wind of QGP flow, leading to smallest observed  hadronization temperature.

\section{Hadronization  at SPS and  at RHIC}\label{hadSR}
\subsection{Method of analysis}\label{Method}
We use the statistical hadronization model presented in 
section \ref{StatHad} to fit experimental particle yields 
and to evaluate the  hadronization conditions including 
the specific strangeness, energy and entropy content.
Experimental results are now  available for several
 collision energies both at SPS and RHIC. Thus it
is possible to study the energy dependence systematically \cite{Raf03}.
We will  see if there is consistency in the fitted energy dependence
both within the energy range  of SPS and RHIC, and between these two 
different energy domains.
Specifically, we reevaluate here our comprehensive analysis of RHIC-130  
Au--Au  collisions \cite{Raf02ga,Raf01kc}, carry out a RHIC-200
preliminary analysis and compare  with 
 SPS Pb--Pb stationary target reactions at 
$\sqrt{s_{NN}^{CM}}=8.75, 12.25,17.2$~GeV  
(projectile energy 40, 80, 158$A$ GeV). 

For SPS, we use here the CERN-SPS-NA49 $4\pi$ particle 
multiplicity  results~\cite{Gaz03a,Gaz03b}, which include 
$\pi^\pm,\ K,\ \overline{K},\ \Lambda,\ \overline{\Lambda},\ \phi$ at
40, 80, 158$A$ GeV. We also  fit (relative) yields  of
 $\Xi,\ \overline\Xi,\ \Omega,\ \overline\Omega$ 
when available. 
Our current results  are found  quite near
to our earlier work reported for
the 158$A$ GeV energy \cite{Raf01kc}, even though  our data sample 
has been limited to (a subset of) NA49 $4\pi$-results, while
our earlier work was dominated by WA97 central rapidity results. 
Since we fit only $4\pi$-particle yield, no information about
the collective flow velocity is obtained. 

Our statistical
hadronization program  allows for decays of resonances, and 
for non-equilibrium occupancies of light quarks $\gamma_q\ne 1$ and strange
quarks $\gamma_s\ne 1$. In the results, we discuss here, we enforce 
strangeness conservation. We assume that 50\% of weak decays from $\Xi$ 
to $\Lambda$ and from $\Omega$ to $\Xi$ are inadvertently included in the
yields when these had not been corrected for such decays. 
Particle yields containing this type of {c}ontamination from a 
weak interaction {c}ascade are marked by a subscript `{c}'.
We assume that the 
pions from such weak decays are {\it not} included in the experimental yields
as these pions tracks clearly do not originate in the interaction vertex.

Not all pion producing resonance decays are known, and 
 many have to be guessed given the present status of  the
particle data tables. We have estimated  comparing to expectations based on
an exponentially rising mass spectrum that pion yields may be 
undercounted in the data based 
statistical hadronization programs \cite{Let02gp}. 
However,  one can also argue that pion yield is overcounted: given the sudden
nature of the hadron fireball breakup, microscopic
reactions cannot exactly follow through to  populate the  heavy resonances. 
To estimate the potential for systematic error we study 
the hadronization with pion  yield artificially  changed by $\pm15$\%.
We find, not surprisingly, that there is an anti-correlation to fitted temperature
in the range of $\mp5$\%, respectively, but other statistical parameters
and physical properties remain  stable. 

Some of the particle ratios we consider can be formed from other results
 already considered. However, each measurement has to be considered as being independent and 
often the data which theoretically is related by a product of ratios, 
is in praxis barely consistent.
The usual recommended  procedure which we follow here 
is to include  all the available {\it different} particle ratios in the 
fit,  and to include for each measured data point 
its error in the study of statistical significance. When
several experimental results are available for the {\it same} ratio 
we fit the average result.

\subsection{RHIC analysis}\label{RHICan}
In the study of particle yields at RHIC, we have allowed for 
 both statistical and systematic errors in the experimental data set. 
We show the results of the fit procedure at 
 RHIC in \rt{TmuRHIC}. The top line gives the reaction energy. 
The fitted statistical
parameters follow,  with errors being both statistical and systematic, derived from
variation of the pion yield. The equilibrium
 results for statistical parameters and particle multiplicities 
agree very well with those presented for the case of chemical equilibrium by 
the Krak\'ow group \cite{Bar03}.

\begin{table}[htb]
\caption{
\label{TmuRHIC} The RHIC chemical freeze-out  statistical parameters found for 
 non-equilibrium (left) and semi-equilibrium (right) 
fits to RHIC results. We show $\sqrt{s_{NN}}$,  
the  temperature $T$,  baryochemical potential $\mu_b$, 
strange quark chemical potential $\mu_s$,
strangeness chemical potential $\mu_{\rm S}$, 
the quark occupancy parameters $\gamma_q$ and
$\gamma_s/\gamma_q$, and in the bottom line 
the statistical significance of the fit. 
The star  (*) indicates that there is an upper limit on the value of 
$\gamma_q^2<e^{m_\pi/T}$ (on left), and/or that the value is set (on right).
}\vspace*{0.2cm}
\begin{center}
\begin{tabular}{l|cc|cc}
\hline\hline
$\sqrt{s_{NN}}$\,[GeV]    &200          & 130        &200          & 130    \\[0.1cm]
\hline
$T$\,[MeV]                &$143\pm7$    & $144\pm3$  &$160\pm8$    & $160\pm4$\\[0.1cm]
$\mu_b$\,[MeV]            &$21.5\pm31$&$29.2\pm4.5$   &$24.5\pm3$   &$31.4\pm4.5$ \\[0.1cm]
$\mu_s$\,[MeV]            &$2.5\pm0.2$ &$ 3.1\pm0.2$ &$2.9\pm0.2$ &$ 3.6\pm0.2$ \\[0.1cm]
$\mu_{\rm S}$\,[MeV]            &$4.7\pm0.4$ &$ 6.6\pm0.4$ &$5.3\pm0.4$ &$ 6.9\pm0.4$ \\[0.1cm]
\hline
$\gamma_q$                &$1.6\pm0.3^*$&$1.6\pm0.2^*$& $1^*$ & $1^*$ \\[0.1cm]
$\gamma_s/\gamma_q$       &$1.2\pm0.15$ & $1.3\pm0.1$ &$1.0\pm0.1$&$1.13\pm0.06$ \\[0.1cm]
\hline
$\chi^2/$dof               &2.9/6        & 15.8/24    &4.5/7       & 32.2/25  \\[0.1cm]
$P_{\rm true}$              &90\%+        & 95\%+    &65\%       & 15\%  \\[-0.7cm]
\end{tabular}
\end{center}
 \end{table}

As the bottom section in \rt{TmuRHIC} shows, the  statistical significance is
very good for chemical non-equilibrium fit, and also, we note
that total $\chi^2$ is half as large as it is for the equilibrium case on the right. 
Only for a very large number of degrees of freedom 
one converges toward $\chi^2/$dof=1 when the experimental errors
are correctly evaluated. For a relatively small number of degrees of freedom, 
as is the case in particular  for RHIC-200, we must have 
$\chi^2/$dof significantly smaller than 
unity to believe in the physical significance of the result. The bottom
line presents an estimate of the probability $P_{\rm true}$ using 
the values of $\chi^2$ and dof that the
fit is an appropriate physics model description. 
Since the systematic errors are here taken into account, this is 
an appropriate time to look at this value. Clearly as more data 
for RHIC-200 becomes available, and errors shrink, this evaluation will change.

\begin{table}[htb]
\caption{\label{table3}}
\begin{center}
\begin{tabular}{l|c|c|c}
\hline\hline
\small
{{ratio}}&{RHIC-130}&{non-eq fit}&
{$\chi^2$}\\[0.1cm]
\hline
$\pi^+/ p$               &$9.5\pm 1.4$    &  9.07       & 1.15\\
$\pi^-/\bar p$           &$13.4\pm 0.9$   &  13.15      & 0.08\\
$\bar p/h^- $            &   --             &  0.0459    &-- \\
$\Lambda_c/h^- $         & $0.059\pm0.004 $ &0.0509    & 4.11\\
$\overline{\Lambda_c}/h^-$&$0.042\pm0.004$ &0.0379     & 1.04\\
$\Xi_c^-/h^- $           &$0.0079\pm0.0012$&0.00805     & 0.01\\
$\overline{\Xi_c^-}/h^- $&$0.0066\pm0.001 $& 0.00645   & 0.02\\
$\Omega/h^-$          &$(12\pm 5)10^{-4}$ & $13.2\,10^{-4}$&0.06\\
$(\overline\Omega+\Omega)/h^-$  
                    &$(22\pm 6.5)10^{-4}$ &$24.8\,10^{-4}$&0.19\\
$\Lambda_c/ p   $        & $0.90\pm0.12 $ &   0.747     & 1.63\\
$\overline{\Lambda_c}/\bar p $
                         & $0.93\pm0.19 $ &   0.826     & 0.30\\
$\Xi^-/\Lambda $         & $0.193\pm0.03$ &   0.189     & 0.02\\
$\overline{\Xi^-}/\overline\Lambda $
                         &$0.219\pm0.035$ &   0.207     & 0.12 \\
$\Omega/\Xi^- $          &    --            &   0.164     & --     \\
$\overline\Omega/\overline{\Xi^-}  $
                         &   --             &   0.180     & --    \\
$(\overline\Omega+\Omega)/(\overline{\Xi}+\Xi)$ 
                         & $0.150\pm 0.04$&   0.171     & 0.28\\
$\bar p/p$               &$0.71\pm 0.06$  &   0.674     & 0.36\\
$\overline{\Lambda_c}/\Lambda_c $
                         & $0.71\pm 0.04 $&   0.745     & 0.78\\
$\overline{\Xi}/\Xi $    & $0.83\pm 0.08 $&   0.801     & 0.13\\
$\overline\Omega/\Omega$ &  $ 0.95\pm 0.1$&   0.878     & 0.51\\
$K^+/\pi^+$              & $0.17\pm 0.02$ &   0.195     & 1.59\\
$K^-/\pi^-$              & $0.15\pm 0.02$ &   0.180     & 2.28\\
$K^-/K^+$                &$0.87\pm 0.07$  &   0.923     & 0.57\\
$K^{*0}/K^-$             &$0.26\pm 0.08$  & 0.231       & 0.13\\
$\phi/h^-$               &$0.02\pm 0.002$ &  0.0212     & 0.37\\
$\phi/K^- $              &$0.15\pm 0.03$  &   0.148     & 0.00\\
$\phi/K^{*0}$            &$0.595\pm0.24$  & 0.639       & 0.03\\[-0.4cm]
\end{tabular}
\end{center}
\end{table}

One is often interested to see how well the particle yields  fit
the experiment. 
For RHIC-130, we  obtain the fitted yield ratios of the following 24 results,
along with a few (three) results one would wish to also have available 
(see table \ref{table3}). It is important that
the reader realizes that these ratios can depend significantly on the assumptions
made about the cascade of particles that ensues hadronization, and which we
cannot fully characterize given  lack of experimental data. We have `played' around
with branching ratios of high mass resonances and believe that the changes that
ensue are all within the errors given for the statistical parameters. However, the
individual yields presented below may change, and thus other groups can derive 
slightly different ratios for same statistical set of parameters.

For RHIC-200 we
employ the following 10 relative  yields available  at this time 
(see table \ref{table4}). We see that the presented non-equilibrium multiplicity 
  fits are  good, and the individual contributions to $\chi^2$ are 
satisfactory.

\begin{table}[htb]
\caption{\label{table4}}
\begin{center}
\begin{tabular}{l|c|c|c}
\hline\hline
{{ratio}}&{RHIC-200}&{non-eq fit}&
{$\chi^2$}\\[0.1cm]\hline
$\pi^+/ p$               & --    &  9.86       & -- \\
$\pi^-/\bar p$           &$12.5\pm 1.7$    &  13.0       & 0.09\\
$\bar p/h^- $            &   --             &  0.04759    &-- \\
$\Lambda_c/h^- $         & -- &0.0441    & --\\
$\overline{\Lambda_c}/h^-$&-- &0.0354     & --\\
$\Xi_c^-/h^- $           & -- &0.00657     & --\\
$\overline{\Xi_c^-}/h^- $& -- & 0.00555   & --\\
$\Omega/h^-$            &$(8.9\pm 2)10^{-4}$ &$ 10\,10^{-4}$& 0.31\\
$\Lambda_c/ p $        & -- &   0.687     & --\\
$\overline{\Lambda_c}/\bar p $
                         &-- &   0.738     & --\\
$\Xi^-/\Lambda $         & --  &   0.175     & --\\
$\overline{\Xi^-}/\overline\Lambda $
                         &-- &   0.187     & -- \\
$\Omega/\Xi^- $          &    --            &   0.152     & --     \\
$\overline\Omega/\overline{\Xi^-}  $
                         &   --             &   0.162     & --    \\
$\bar p/p$               &$0.74\pm 0.04$   &   0.746     & 0.02\\
$\overline{\Lambda_c}/\Lambda_c $
                         & --&   0.801     & --\\
$\overline{\Xi}/\Xi $    & -- &   0.844     & -- \\
$\overline\Omega/\Omega$ &   $ 1.05\pm 0.2$&   0.902     & 0.55\\
$K^+/\pi^+$              & -- &   0.182     & --\\
$K^-/\pi^-$              & $0.156\pm 0.02$ &   0.172     & 0.63\\
$K^-/K^+$                &$0.95\pm 0.05$   &   0.945     & 0.01\\
$K^{*0}/K^-$             &$0.205\pm 0.033$ & 0.227       & 0.44\\
$\phi/h^-$               &$0.02\pm 0.002$  & 0.0183      & 0.74\\
$\phi/K^- $              &$0.118\pm 0.04$  &   0.133     & 0.14\\
$\phi/K^{*0}$            &$0.595\pm0.123$  & 0.585       & 0.01\\[-0.4cm]
\end{tabular}
\end{center}
\end{table}

The quality of RHIC results presented  in \rt{TmuRHIC}   confirms that  
the  statistical hadronization mechanism is an appropriate particle
production model in presence of the {\em sudden QGP
breakup} which without doubt is seen at RHIC.  However, 
 the statistical hadronization approach 
may be  not entirely appropriate at low SPS energies, and 
all AGS energies. It is possible that 
kinetic models need to  be applied, which  allow 
for a multitude of freeze-out conditions
depending on the nature of particle considered. For this reason, when  
we study this energy domain we will look more closely at the 
behavior of individual observable ($K/\pi$) rather than at the global 
yields of all particles, the method we are using at RHIC and high SPS 
energies. 

\subsection{Phase space occupancy, energy stopping as function of 
collision energy}\label{StatPar}
The non-equilibrium parameters $\gamma_s,\gamma_q$ 
we find in our fits are shown  in \rf{PLGAQS} as function of collision energy.
In top figure, open squares show the result 
when we fix $\gamma_q=1$. We note that in this case  the SPS fits 
 yield  $\gamma_s\simeq 0.7$.  However,  we
see that $\gamma_q>1$ is preferred in all cases, with value converging toward
the limit $\gamma_q\to \gamma_q^{\rm max}=e^{m_\pi/2T}$. This behavior reflects on 
the need to account for the  excess in charged hadron multiplicity. 
With this large $\gamma_q$ also $\gamma_s$ is found to be quite large,
in particular so at RHIC. We need to solidify  this result since this 
is indeed the expected signature of the sudden hadronization of deconfined 
strangeness rich QGP phase.

\begin{figure}[htb]
\hspace*{1.5cm}\psfig{width=8.6cm,figure=\pathnow 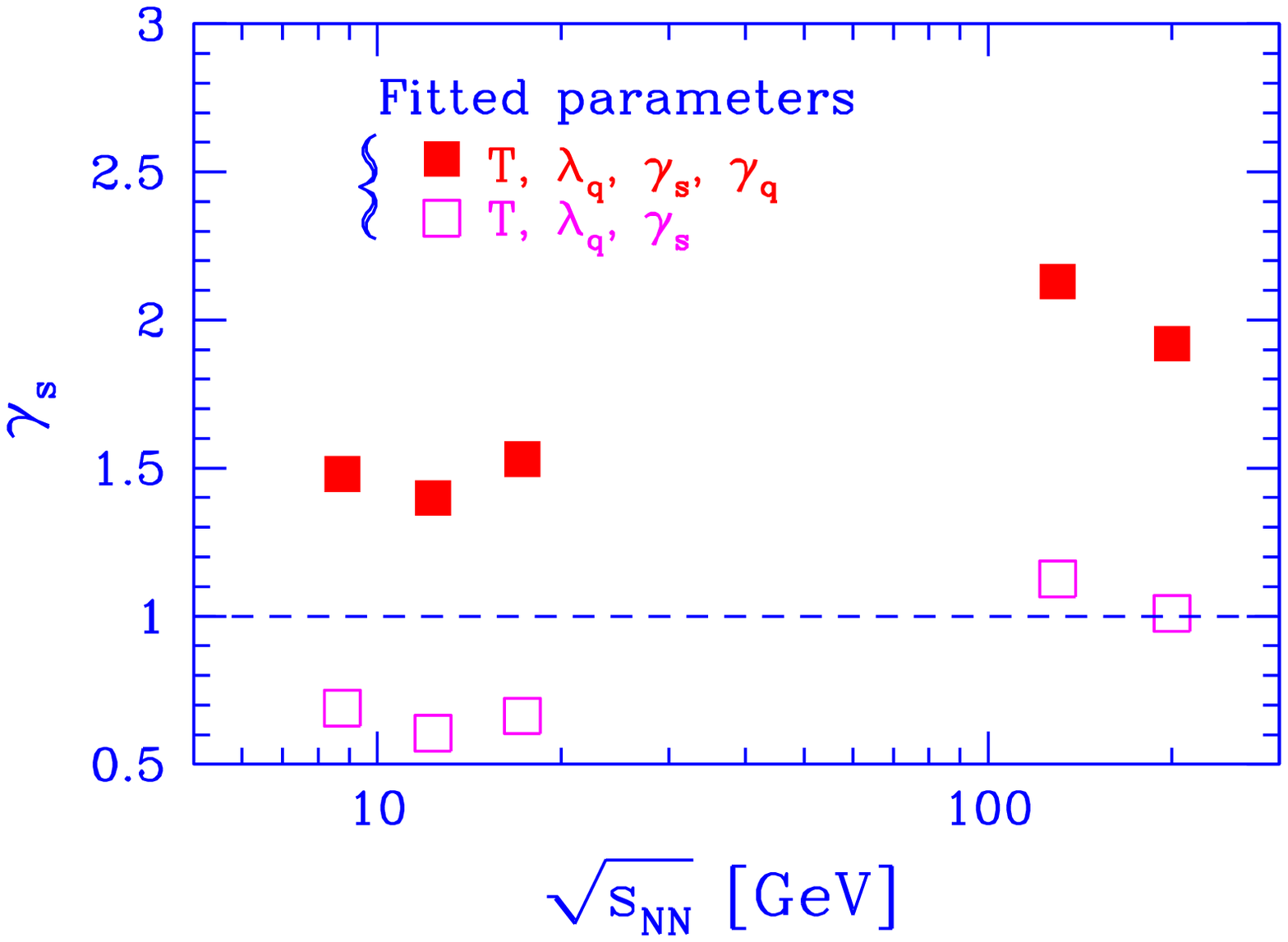}\\
\hspace*{1.5cm}\psfig{width=8.6cm,figure=\pathnow 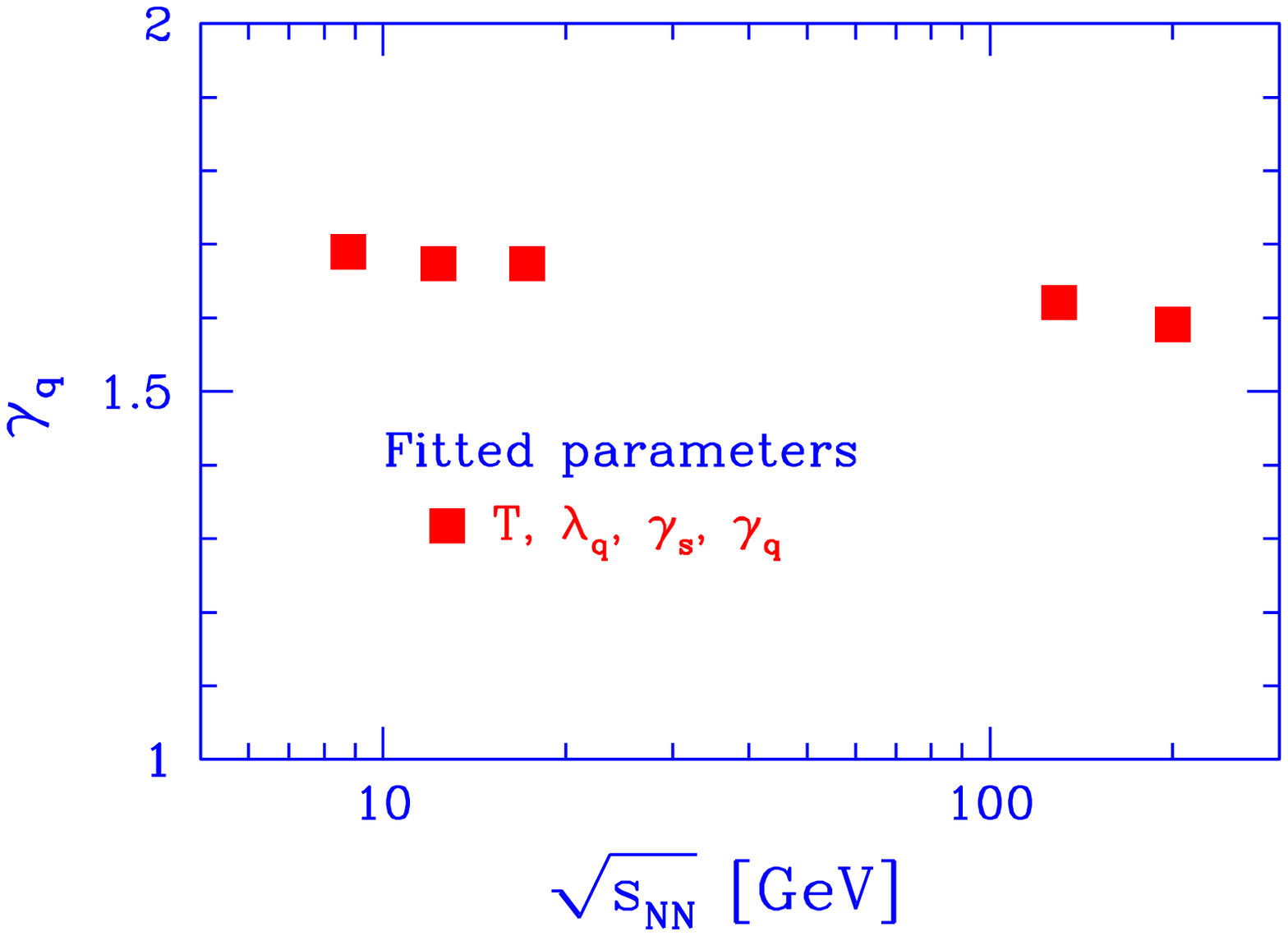}
\caption{\label{PLGAQS}
Fits of, top to  bottom 
$\gamma_s$ (top) and $\gamma_q$ (bottom)
chemical non-equilibrium parameters as function of collision 
energy. Open squares results for $\gamma_q=1$.
}
\end{figure} 

We next  address the reaction mechanisms, evaluating the  energy 
stopping. We define it  as  the ratio of the thermal energy 
$E^{\rm th}_{{\rm i}\,NN}$ found in particles produced 
and normalized to a nucleon pair, with the  
collision energy $\sqrt{s_{NN}}$ which is also given per 
colliding nucleon pair. This study  will
allow us to consider the energy dependence of other 
variables  as function of the energy available $E^{\rm th}_{{\rm i}\,NN}$
in the fireball,  rather than as function of the energy of 
colliding nuclei $\sqrt{s_{NN}}$, which
in its greatest part may not participate in the reaction. 

The open triangles \rf{PLEBS} represent energy 
stopping for the fit assuming
full chemical equilibrium. We note that at SPS this leads to 
the absurd behavior that stopping is increasing with increasing
collision energy, reaching 70\% at the top energy. This alone
proves  that the chemical equilibrium  approach to statistical 
hadronization is physically meaningless.

\begin{figure}[t]
\hspace*{1.2cm}\psfig{width=9.6cm,figure=\pathnow  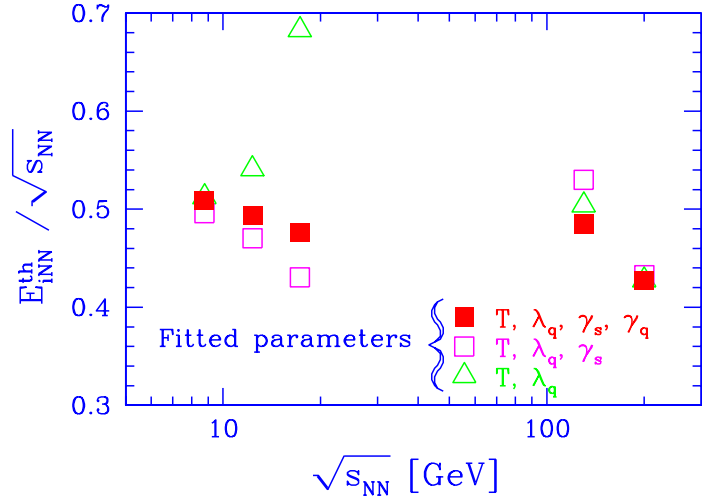}
\caption{\label{PLEBS}
Fraction of energy stopping at SPS and RHIC: 
results are shown for 40, 80 and 158$A$ GeV Pb--Pb fixed target SPS reactions 
and for  130 and 200$A$ GeV Au--Au RHIC  interactions. 
}
\end{figure}

The open squares in \rf{PLEBS} show the stopping found for 
the case of semi-equilibrium fit,
{\it i.e.},  allowing only strange quarks to be out of chemical equilibrium.
The full squares  in \rf{PLEBS} apply full chemical non-equilibrium with phase space
occupancy shown in \rf{PLGAQS}. Up to the step-up between SPS and RHIC
we see a smooth fall of of the stopping with increasing collision 
energy when chemical non-equilibrium
is allowed for. We recall that the statistical hadronization
study we have made assumes tacitly that Bjorken rapidity  scaling applies exactly.
One would hope  that the RHIC stopping result will  come down when we allow for 
the non-scaling yields of particles.

\subsection{Strangeness and entropy  in statistical hadronization}\label{StrangeS}
We would like to compare the efficiency of strangeness and entropy 
production between the two widely different systems considered.  
We  evaluate the number of strange quark pairs 
produced (strangeness  yield) and divide it by the computed 
thermal fireball  baryon number. This eliminates the need for an
absolute  yield normalization. More generally, a ratio of two extensive computed
variables is nearly independent of the  dynamics of hadronization.

In   \rf{PLETRBSLOG}, we present  strangeness
production per baryon. The top \rf{PLETRBSLOG} section 
 shows our result as a function of the collision energy. 
When using the thermal energy content in the reaction, we 
arrive at the result with a smoothly rising curve seen in the bottom figure
 section.  As noted in the  insert, strangeness yield is 
 rising faster than linearly with fireball energy content.

\begin{figure}[ht]
\hspace*{1.90cm}\psfig{width=7.5cm,figure=\pathnow  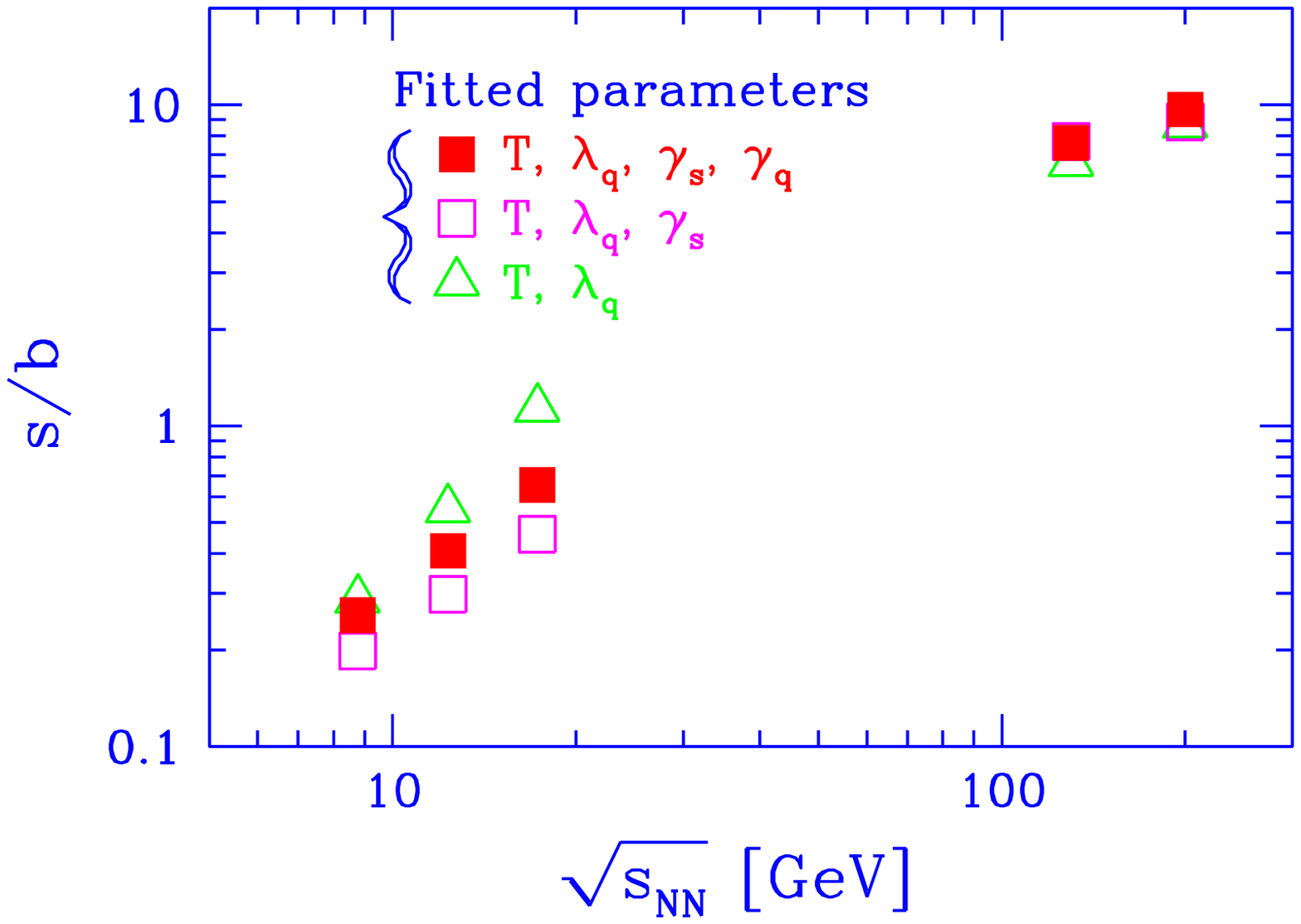}\\
\hspace*{1.90cm}\psfig{width=7.5cm,figure=\pathnow  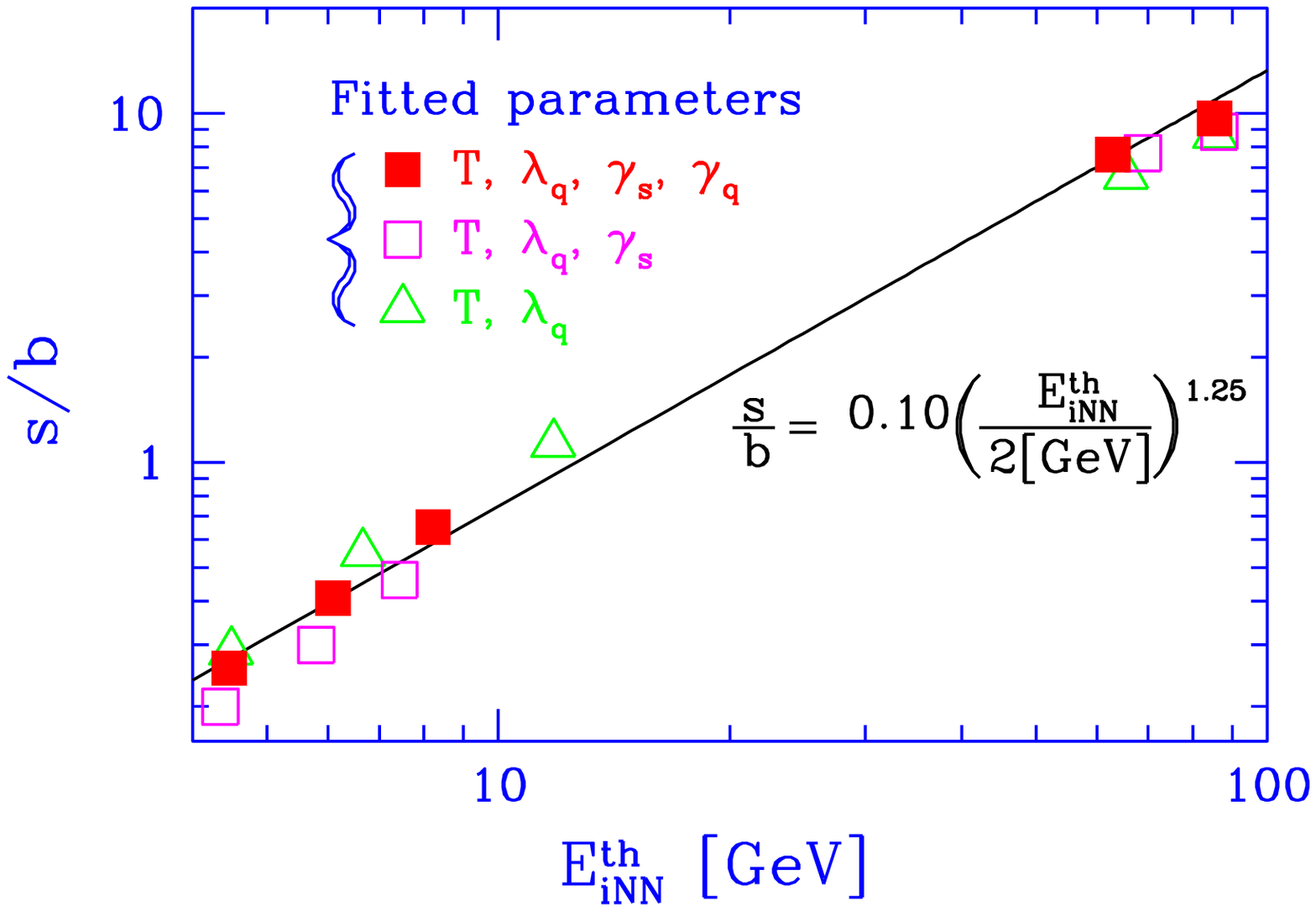}
\caption{\label{PLETRBSLOG}
Strangeness per thermal  baryon  participant, $s/b$  as function of 
 $\sqrt{s_{NN}}$ (top), and  as function of thermal specific energy content
$E^{\rm th}_{{\rm i}\,NN}$ (bottom). 
}
\end{figure}

We next  look at
strangeness per entropy,\ $s/S$,  shown in \rf{PLETRSEIC}. There is 
considerable physical importance of  $s/S$, since both entropy $S$
and strangeness $s$ are produced early on in kinetic processes,
are nearly conserved in the  hydrodynamical expansion of QGP, and 
increase only moderately in the hadronization of QGP. 
Thus the value seen for $s/S$ is reflecting on the 
kinetic mechanisms operational in early stages of the heavy ion 
collision. Because there is strangeness mass as energy threshold to overcome, 
one would expect that as the initial QGP phase becomes hotter with 
increasing collision energy, the excitation of strangeness grows
more rapidly than excitation of entropy.

\begin{figure}[ht]
\hspace*{1.20cm}\psfig{width=9.cm,figure=\pathnow  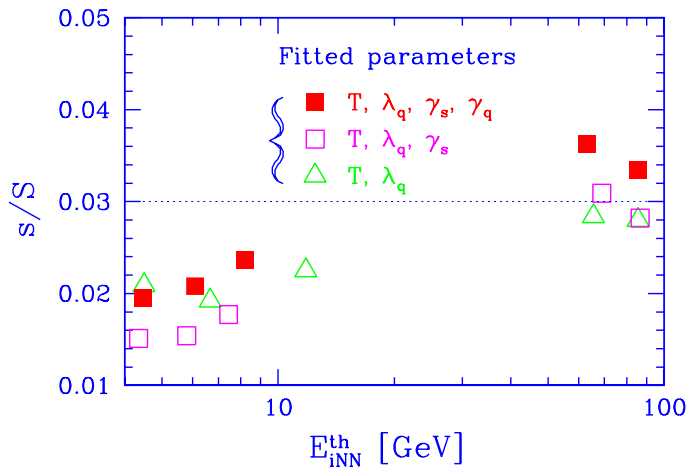}
\caption{\label{PLETRSEIC}
Strangeness per  
entropy $s/S$ as function of $E^{\rm th}_{{\rm i}\,NN}$.
}
\end{figure}

We note, in \rf{PLETRSEIC}, for the chemical non-equilibrium assumption,
 a smooth and slow increase of $s/S$, 
from 0.02 to 0.025 at SPS. The RHIC-130 point is within the smooth extrapolation
of the SPS result. The low value of RHIC-200 result is probably expressing 
a systematic error related to the absence of multistrange particles in the
fit. While entropy is already fully accounted for, only availability of 
multistrange hadron results will allow to obtain a reliable strangeness
yield.  However, if this expectation is not fulfilled, it will be
very important to obtain experimental results in the energy range
between SPS and  RHIC,  in the domain where the change of the systematic behavior 
(slope) of $s/S$ as function of energy occurs.

\section{Search for an  Energy Threshold}\label{compare}
\subsection{Kaon to pion ratio}\label{Kpi}
One of the most interesting questions is if there is 
an  energy threshold for the formation of a new state of matter. 
In our study of strangeness per entropy $s/S$, in subsection \ref{StrangeS},
we have  seen a possibility that new physics sets in between the SPS and
 RHIC-130 energy domain. However, a final verdict on this issue depends 
on  further  RHIC-200 results. 

The NA49 experiment has been exploring 
the low energy domain at SPS in search for 
a new physics threshold between AGS and SPS energy range. 
We refer to the report
by Marek Ga\'zdzicki for the set of  arguments
that just at the bottom of SPS energy range
something new happens \cite{Marek}.
Perhaps the most important evidence cited by Marek is 
a peak in the $K^+/\pi^+$ ratio, see top section 
of \rf{Kpisqrt} \cite{Afa2002mx}. 

\begin{figure}[ht]
\vskip-.1cm
\hspace*{1.9cm}\psfig{width=8.2cm,height=8.8cm,figure=\pathnow  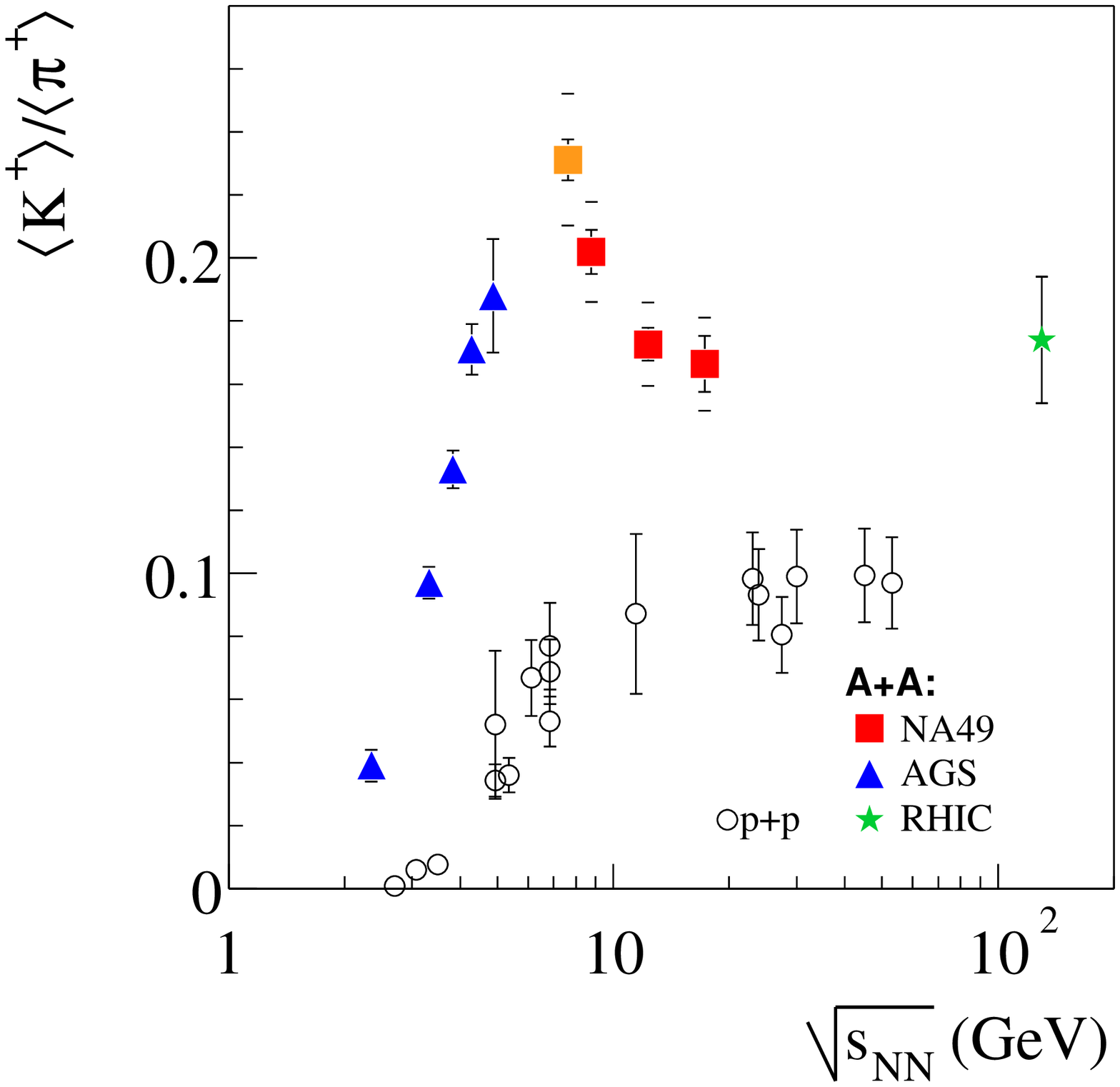}
\vskip-.1cm
\hspace*{2.4cm}\psfig{width=7.8cm,height=6.5cm, figure=\pathnow  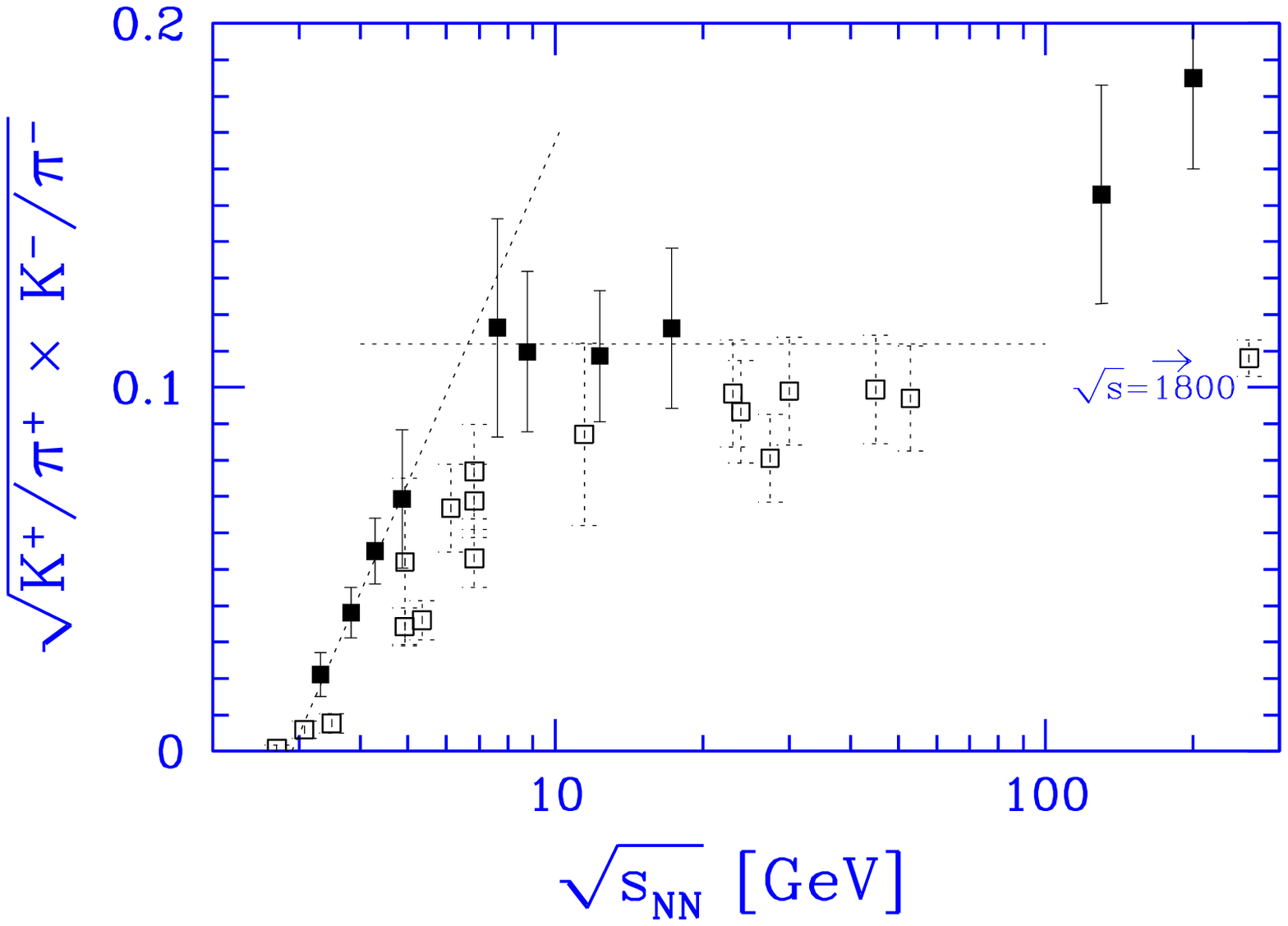}
\vskip-.1cm
\caption{\label{Kpisqrt}
Top: $4\pi$-ratio of yields of  $K^+/\pi^+$
for nuclear (filled symbols) and elementary interactions (open symbols);
(courtesy of NA49 collaboration \protect\cite{Gaz03b,Lee03}).
Bottom: the  reduced 
$K/\pi$ ratio for nuclear (filled) and elementary collisions (open symbols)
both as function of collision energy. 
Figures rescaled to the same magnitude on abscissa and 
ordinate.}
\end{figure}

We think that this peak is purely an effect of the dense baryon medium. 
As the collision energy increases, beginning with low energies,
the  baryon density increases. However, at some   collision energy 
baryons shoot through and the baryon density begins to drop. Since there is 
considerable sensitivity to baryon chemical potential and thus baryon 
density in the $K^+/\pi^+$ ratio considered
by the NA49 collaboration,  we prefer to consider the nearly 
baryon density  independent ratio \req{K2pi}, 
\beql{Kdpi}
\frac{K}{\pi}=\sqrt{  \frac{K^+}{\pi^+}\, \frac{K^-}{\pi^-}}.
\eeq

We present ${K}/{\pi}$ double ratio 
in the bottom of \rf{Kpisqrt}  as black filled square.
At high energy, we show the two RHIC results and at SPS
and AGS energies we use the NA49 data set \cite{Afa2002mx,Lee03}.
The open squares are for the charged ratio K$^+/\pi^+$ 
measured in  $pp$ reactions, which is offering an upper limit on 
${K}/{\pi}$ from $pp$ reactions, also using NA49 data set. We indicate for 
$\sqrt{s}={1800}$ the $p$--$\bar p$ TEVATRON result \cite{Ale93wt}.

Both upper and lower portion of  \rf{Kpisqrt} are drawn on same scale.
 Comparing the top and bottom in  \rf{Kpisqrt}, we see that the
peak between AGS and SPS energy domain has completely disappeared, and that
in the SPS energy domain the $K/\pi$ ratio \req{Kdpi} is flat.
This behavior we already saw in the ratio of strangeness to entropy, bottom
of \rf{PLETRBSLOG}. Moreover, when we look at the inverse slopes of
the K$^+$ $m_\bot$-distributions, \rf{GazTKpnew}, we again see 
in SPS domain a flat distribution \cite{Gaz03b}. 

\begin{figure}[ht]
\vspace*{0cm}
\hspace*{2cm}
\psfig{height=7cm,figure=\pathnow 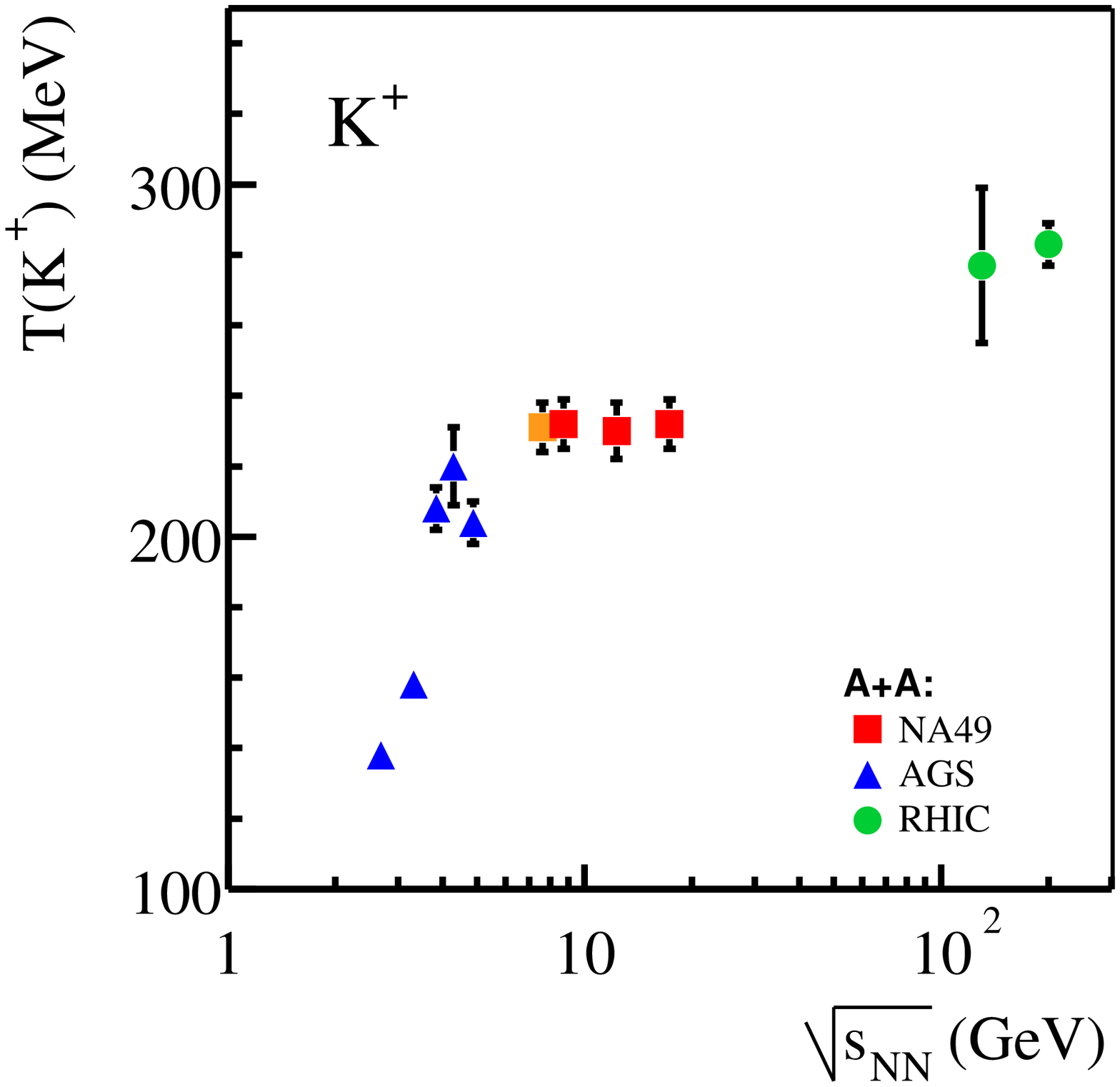}
\vspace*{0cm}
\caption{ 
The inverse slope $T$ of K$^+$ spectra  at AGS (filled triangles) 
at SPS (filled squares, including a recent 30$A$ GeV NA49 experiment and
at RHIC-130 and RHIC-200.
\label{GazTKpnew}}
\end{figure}

Although the large error bars leave a lot of space for further structure,
we do not see any evidence for a `peak' in strangeness production,
rather we believe that NA49 has discovered  that $K^+/\pi^+$ ratio  allows
to determine where the baryon density peaks as function of collision
energy.  

As the dashed lines in the bottom portion of 
 \rf{Kpisqrt} suggest, there is a smooth transition
from rising to saturated behavior of ${K}/{\pi}$  ratio. However
this behavior mirrors the result we see in the $pp$ reactions, except that
it occurs at lower collision energy. {\em We conclude that if kaon and pion yields are
solely used as the measuring stick for new physics, there is clear evidence
 that new physics  is seen at RHIC, where a true deviation from $pp$, $p\bar p$ 
behavior occurs.}

\subsection{A `conventional' $K/\pi$ excess}\label{LHC}
The enhancement of $K/\pi$ ratio at RHIC may be, however, simply 
result of novel mechanism of  pion suppression. We first recall that
there is considerable increase in the transverse velocity of matter flow, as
is born out, {\it e.g.}, in the higher inverse slopes of particle  
spectra at RHIC. We see this effect in  \rf{GazTKpnew}  for  $K^+$. 

The kinetic energy of this
transverse motion must be taken from the thermal energy of the expanding matter,
and ultimately this leads to local cooling and thus a reduction in the number of 
quarks and gluons. The local entropy density decreases, but the expansion
assures that the total entropy still increases. Primarily, gluons are
feeding the expansion dynamics, while strangeness being  weaker
coupled remain least influenced by this dynamics. Model calculations 
show that there is practically no strangeness reannihilation in transverse
expansion at RHIC \cite{Raf99R}. 

The depletion of the non-strange degrees
of freedom in the feeding of the expansion assures an increase in the 
$K/\pi$ ratio with increase of collision energy.
This effect explains  why
$\gamma_s>\gamma_q$, see table \ref{TmuRHIC}. If depletion by  
transverse expansion dynamics of the
non-strange hadrons is the origin of the increase in $K/\pi$ at RHIC, 
then this effect would arise gradually as function of collision 
energy. If, however, new phase of matter is the cause for this increase, we 
would expect a more `edgy'  onset of the increase in the energy domain between 
SPS and RHIC. 

An even greater effect can be expected at future CERN LHC collider. 
for heavy ions LHC will be a 6000GeV collider, 30 times the maximum energy at RHIC.
The transverse expansion dynamics effects should be much greater at LHC
and we expect  significantly grater  $K/\pi$ ratio, and thus 
greater value of $\gamma_s/\gamma_q$. The only limit that we can
see for a rise in $\gamma_s/\gamma_q$, which is the factor controlling the rise 
in  $K/\pi$, is the condensation limit, \req{Kcond}:
\[
{\rm since}\ \gamma_q^2\ \le\ e^{m_\pi\over T}\ {\rm and}\  \gamma_s\gamma_q \ \le\ 
e^{m_K\over T} \ {\rm we\ have}\
\to\  
\gamma_s/\gamma_q\ \le\  e^{{m_K-m_\pi}\over T}.
\]
We have shown the resulting maximal value of $K/\pi$ in \rf{KPIT} as a solid line
as function of hadronization temperature. 

\begin{figure}[ht]
\vspace*{-4.7cm}
\centerline{\epsfig{width=12.cm,clip=,figure=\pathnow 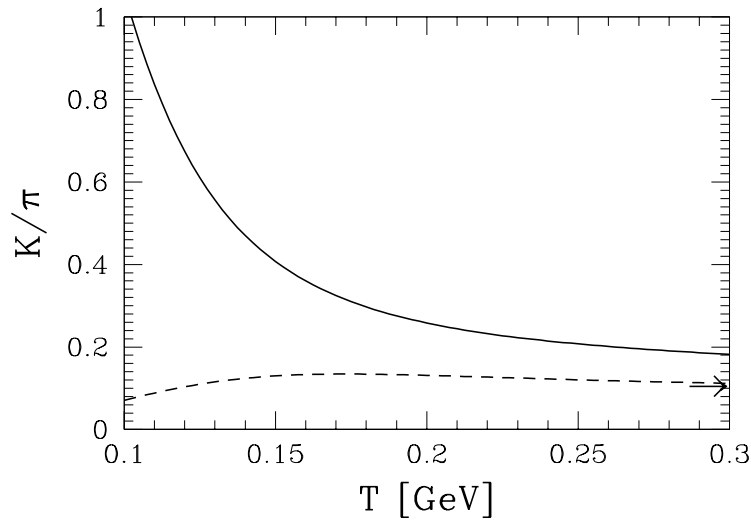}
}
\vspace*{-0.5cm}
\caption{ 
Kaon to pion ratio $K/\pi$ (see text) as function of 
hadronization temperature $T$ for chemical equilibrium (dashed line)
and with maximum allowable  $\gamma_s$ and $\gamma_q$ (solid line).
\label{KPIT} 
}
\end{figure}

The dashed line  in \rf{KPIT} shows the result when $\gamma_s=\gamma_q=1$. 
Depending on what we believe to be a valid hadronization temperature for
a fast transversely expanding fireball, the possible enhancement in the 
$K/\pi$ ratio will be in the range of a factor 2--3. The arrow
indicates whereto the ratio converges for $T\to \infty$. This limit 
applies to both solid and dashed lines and critically depends on
the high mass resonances which we really do not know very well. 

It can be assumed that  hadronization at LHC occurs
in a temperature domain $T=160\pm 20$\,MeV. In this case
 even more interesting than the $K/\pi$ ratio enhancement would be the associated 
enhancement anomaly in strange (antibaryon) yields. One would find a strong 
inversion in the population, with the more  strange  baryons and antibaryons
 being more abundant than less strange species. 

\subsection{Final remarks}\label{final}
We have shown that strangeness 
at CERN-SPS and BNL-RHIC is a well developed  tool
allowing  to  study   QGP-hadronization. Our 
discussion has shown that a systematic study 
of strange hadrons  fingerprints the properties 
of  a new state of matter. We have argued that the
deconfinement specific observable 
is strangeness,  and we have explained why we 
have little doubt that deconfined phase  has been 
formed in relativistic heavy ion collisions.

The  situation about deconfinement  onset as function of
energy is today  much less clear than ever
before. We argue here (subsection \ref{Kpi}) 
that between AGS and SPS energy  range
a smooth transition to a saturated
$K/\pi$ yield is seen. This behavior mirrors the behavior of slopes 
of $m_\bot$ kaon
spectra. At SPS, we note a great enhancement of strange
antibaryon yields, rising with strangeness content. Analysis of 
strange hadrons shows that the specific strangeness
yield is very high. The strangeness production and hadronization
at SPS is consistent with the behavior of the QGP phase. 

Somewhere between SPS and RHIC the yield of 
$K/\pi$ begins to rise above the elementary reaction result.
We saw, in the same energy range, a possible 
anomaly in the strangeness to entropy ratio $s/S$, subsection \ref{StrangeS}. 
However, we argue in subsection \ref{LHC} that the $K/\pi$ 
ratio rise could be a dynamical effect due to rapid transverse expansion, 
and we discuss in subsection \ref{StrangeS} that 
 the $s/S$ anomaly could be removed by more comprehensive results 
forthcoming from RHIC-200 run. 
Thus, at this time, RHIC could harbor some great new discovery, or be a QGP
producer with  a more powerful transverse expansion and grater energy density. 
The behavior of particle spectra support this view \cite{Bro02}.

The BNL-RHIC experimentalists announced recently  the discovery of new
physical phenomena, jet quenching,  as would be expected in  QGP. 
In February 2000 CERN-SPS experimental groups have combined their experimental
results  to claim a discovery of new state of matter, which looks like QGP.  
Aside of strange hadron evidence, several other observables were 
implicated  in the CERN announcement. 
Discovery of QGP is also claimed in the study of
$pp$ reactions \cite{Ale02eh}. This work evaluates  the ideal gas
degrees of freedom which number is enhanced as expected in QGP  phase.

Though there
was not much AGS public relation campaign,  we do not
see a reason why deconfined state formation  did not occur at these
low energies. If
the criterion of relevance is the heating of nuclear matter to 
the phase boundary found in  lattice gauge calculations, this
is achieved at the mid-range of AGS  collision energy. However, in
this case the new phase of matter is embedded in the center of a 
opaque hadron shell. Only electromagnetic probes would have allowed to
see into the heart of such low energy  fireball. A new experimental 
facility with
much more intense beams is needed to perform such experiments.

A student reading this report
may find a useful summary of facts about QGP,
physics of strangeness, and statistical hadronization. The 
advanced QGP researcher will find here
practical and useful results addressing  analysis 
of the RHIC-130 and
RHIC-200 experimental data, and a comparison with  SPS results. These results
do suggest that through study of strangeness at RHIC we should soon
understand better the properties of the hot hadron fireball and its
presumably deconfined structure.

\section*{Acknowledgments}
Supported  by a grant from the U.S. Department of
Energy,  DE-FG03-95ER40937\,. Laboratoire de Physique Th\'eorique 
et Hautes Energies, \linebreak LPTHE, at  University Paris 6 and 7 is supported 
by CNRS as Unit\'e Mixte de Recherche, UMR7589.



\end{document}